\documentclass[aos,MSNbibl,seceqn,citesort,dvips]{arximspdf}
\usepackage[section]{algorithm}
\usepackage{algorithmic}
\usepackage{mathbh}
\usepackage{graphicx}

\doi{10.1214/11-AOS940}
\volume{40}
\issue{1}
\pubyear{2012}
\firstpage{294}
\lastpage{321}

\makeatletter

\newcommand{\ddcup}{\, \dot{\cup}\,}

\newtheorem{lemma}{Lemma}[section]
\newtheorem{theo}{Theorem}[section]

\newproclaim{remark}{Remark}[section]
\newproclaim{defi}{Definition}[section]
\newproclaim{example}{Example}

\newcommand{\Prob}{\mathbb{P}}

\newcommand{\ci}{\perp\hspace*{-6pt}\perp}
\newcommand{\newnci}{\mbox{$\not\hspace*{1pt}\perp\hspace*{-11pt}\perp$}}

\newcommand{\leftstar}{{*}{\!\!\rightarrow}}
\newcommand{\rightstar}{{\leftarrow\!\!}{*}}

\newcommand{\leftcircarrow}{
  \setlength{\unitlength}{1mm}
  \begin{picture}(5,1)(0,0)
    \put(1,1){\circle{1}}
    \put(1.2,0){$\rightarrow$}
  \end{picture}
}

\newcommand{\twostar}{
  \setlength{\unitlength}{1mm}
  \begin{picture}(5,1)(0,0)
    \put(0.2,0.1){$*$}
    \put(1.3,1){\line(1,0){2.4}}
    \put(2.8,0.1){$*$}
  \end{picture}
}

\newcommand{\rightstarleftcirc}{
\setlength{\unitlength}{1mm}
\begin{picture}(5,1)(0,0)
\put(1,1){\circle{1}}
\put(1.5,1){\line(1,0){2.4}}
\put(2.9,0.1){$*$}
\end{picture}
}

\newcommand{\leftstarrightcirc}{
\setlength{\unitlength}{1mm}
\begin{picture}(5,1)(0,0)
\put(0.2,0.1){$*$}
\put(1.1, 1){\line(1,0){2.4}}
\put(4, 1){\circle{1}}
\end{picture}
}

\newcommand{\leftcirc}{
\setlength{\unitlength}{1mm}
\begin{picture}(5,1)(0,0)
\put(1,1){\circle{1}}
\put(1.5,1){\line(1,0){3}}
\end{picture}
}

\newcommand{\twocirc}{
\setlength{\unitlength}{1mm}
\begin{picture}(5,1)(0,0)
\put(1,1){\circle{1}}
\put(1.5,1){\line(1,0){2}}
\put(4,1){\circle{1}}
\end{picture}
}

\makeatother

\begin{document}
\begin{frontmatter}

\title{Learning high-dimensional directed acyclic graphs
with latent and selection variables}
\runtitle{Learning DAGs with latent and selection variables}

\begin{aug}
\author[A]{\fnms{Diego} \snm{Colombo}\corref{}\thanksref{t1}\ead[label=e1]{colombo@stat.math.ethz.ch}},
\author[A]{\fnms{Marloes H.} \snm{Maathuis}\thanksref{t1}\ead[label=e2]{maathuis@stat.math.ethz.ch}},
\author[A]{\fnms{Markus} \snm{Kalisch}\ead[label=e3]{kalisch@stat.math.ethz.ch}}
\and
\author[B]{\fnms{Thomas S.} \snm{Richardson}\thanksref{t2}\ead[label=e4]{thomasr@u.washington.edu}}
\runauthor{Colombo, Maathuis, Kalisch and Richardson}
\affiliation{ETH Z\"urich, ETH Z\"urich, ETH Z\"urich and University of
Washington}
\address[A]{D. Colombo\\
M. H. Maathuis\\
M. Kalisch\\
ETH Z\"urich\\
Seminar for Statistics\\
R\"amistrasse 101\\
8092 Z\"urich\\
Switzerland\\
\printead{e1}\\
\phantom{E-mail: }\printead*{e2}\\
\phantom{E-mail: }\printead*{e3}} 
\address[B]{T. S. Richardson\\
Department of Statistics\\
University of Washington\\
Seattle, Washington 98195\\
USA\\
\printead{e4}}
\end{aug}

\thankstext{t1}{Supported in part by Swiss NSF Grant
200021-129972.}
\thankstext{t2}{Supported in part by U.S. NSF Grant CRI 0855230 and
U.S. NIH Grant R01 AI032475.}

\received{\smonth{4} \syear{2011}}
\revised{\smonth{8} \syear{2011}}

%
\begin{abstract}
We consider the problem of learning causal information between random
variables in directed acyclic graphs (DAGs) when allowing arbitrarily
many latent and selection variables. The FCI (Fast Causal Inference)
algorithm has been explicitly designed to infer conditional
independence and causal information in such settings. However, FCI is
computationally infeasible for large graphs. We therefore propose the
new RFCI algorithm, which is much faster than FCI. In some situations
the output of RFCI is slightly less informative, in particular with
respect to conditional independence information. However, we prove that
any causal information in the output of RFCI is correct in the
asymptotic limit. We also define a~class of graphs on which the outputs
of FCI and RFCI are identical. We prove consistency of FCI and RFCI in
sparse high-dimensional settings, and demonstrate in simulations that
the estimation performances of the algorithms are very similar. All
software is implemented in the R-package \texttt{pcalg}.
\end{abstract}

%

\end{frontmatter}

\section{Introduction}

We consider the problem of learning the causal structure between random
variables in acyclic systems with arbitrarily many latent and selection
variables.
As background information, we first discuss the situation without latent
and selection variables in Section~\ref{secintronohidden}.
Next, in Section~\ref{secintrohidden} we discuss complications that
arise when allowing for arbitrarily many latent and selection
variables. Our new contributions are outlined in Section~\ref{secintrocontributions}.

\subsection{Systems without latent and selection variables}
\label{secintronohidden}

We first consider systems that satisfy the assumption of causal
sufficiency, that is,
that there are no unmeasured common causes and no unmeasured selection
variables. We assume that causal information between variables can be
represented by a~directed acyclic graph (DAG) in which the vertices represent random
variables and the edges represent direct causal effects (see, e.g.,
\cite{Pearl00,Pearl09,SpirtesEtAl00}). In particular,
$X_1$ is a~direct cause of $X_2$ only if $X_1 \rightarrow X_2$ (i.e., $X_1$
is a~parent of $X_2$), and $X_1$ is a~(possibly indirect) cause of $X_2$
only if there is a~directed path from $X_1$ to $X_2$ (i.e., $X_1$ is an
ancestor of $X_2$).

Each causal DAG implies a~set of conditional independence relationships
which can be read off from the DAG using a~concept called $d$-separation~\cite{Pearl00}. Several DAGs can describe exactly the same conditional
independence
information. Such DAGs are called Markov equivalent and form a~Markov
equivalence
class. For example, consider DAGs on the variables $\{X_1,X_2,X_3\}$.
Then $X_1
\rightarrow X_2 \rightarrow X_3$, $X_1 \leftarrow X_2 \leftarrow X_3$ and
$X_1 \leftarrow X_2 \rightarrow X_3$ form a~Markov equivalence class,
since they
all imply the single conditional independence relationship $X_1 \ci X_3 |
X_2$, that is, $X_1$ is conditionally independent of $X_3$ given~$X_2$
(using the shorthand notation of Dawid~\cite{Dawid80}). Another Markov
equivalence class is given by
the single DAG $X_1 \rightarrow X_2 \leftarrow X_3$, since this is the only
DAG that implies the conditional independence relationship $X_1 \ci
X_3$ alone. Markov equivalence classes of DAGs can be described
uniquely by a~completed
partially directed acyclic graph (CPDAG) \mbox{\cite{AndersonEtAll97,Chickering02}}.

CPDAGs can be learned from conditional independence
information if one assumes faithfulness, that is, if the conditional
independence relationships among the variables are exactly equal to those
that are implied by the DAG via $d$-separation. For example, suppose that the
distribution of $\{X_1,X_2,X_3\}$ is faithful to an unknown underlying
causal DAG, and that the only conditional independence relationship is $X_1
\ci X_3 | X_2$. Then the corresponding Markov equivalence class
consists of $X_1
\rightarrow X_2 \rightarrow X_3$, $X_1 \leftarrow X_2 \leftarrow X_3$ and
$X_1 \leftarrow X_2 \rightarrow X_3$, and we know that one of
these three DAGs must be the true causal DAG. Algorithms that are based on
this idea are called constraint-based algorithms, and a~prominent example
is the PC algorithm~\cite{SpirtesEtAl00}. The PC algorithm is
sound (i.e., correct) and complete (i.e., maximally informative) under
the assumptions of
causal sufficiency and faithfulness~\cite{SpirtesEtAl00}. It is
efficiently implemented in the R-package \texttt{pcalg}
\cite{KalischEtAl11}, and was shown to be asymptotically
consistent in sparse high-dimensional settings~\cite{KalischBuehlmann07a}.

In practice, one often wants to estimate not only the Markov
equivalence class of
DAGs, but also the size of causal effects between pairs of variables. In
the special case that the estimated CPDAG represents a~single DAG, one
can do this via, for example, Pearl's do-calculus (also called intervention
calculus; see~\cite{Pearl00}) or marginal structural models
\cite{RobinsEtAl00}. If the estimated CPDAG represents
several DAGs, one can conceptually estimate causal effects for each DAG in
the Markov equivalence class, and use these values to infer bounds on causal
effects. This idea, together with a~fast local implementation, forms the
basis of the IDA algorithm~\cite{MaathuisColomboKalischBuhlmann10,MaathuisKalischBuehlmann09} which
estimates bounds on causal effects from observational data that are
generated from an unknown causal DAG (IDA stands for Intervention
calculus when the DAG is Absent).
The IDA algorithm was shown to be consistent in sparse high-dimensional
settings~\cite{MaathuisKalischBuehlmann09}, and was validated on a~challenging
high-dimensional yeast gene expression data set~\cite{MaathuisColomboKalischBuhlmann10}.

\subsection{Complications arising from latent and selection
variables}\label{secintrohidden}

In practice there are often latent variables, that is, variables
that are not measured or recorded. Statistically speaking, these
variables are marginalized out. Moreover, there can be selection
variables, that is,
unmeasured variables that determine whether or not a~measured unit is
included in the data sample. Statistically speaking, these variables are
conditioned on (see~\cite{Cooper95,SpirtesMeekRichardson99} for
a~more detailed discussion). Latent and selection variables cause
several complications.

\begin{figure}[b]
\begin{tabular}{@{}ccc@{}}

\includegraphics{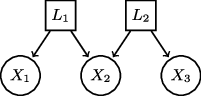}
 & \includegraphics{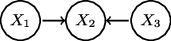} & \includegraphics{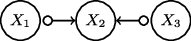}\\
(a) & (b) & (c)
\end{tabular}
\caption{Graphs corresponding to the examples in Section
\protect\ref{secintrohidden}. Throughout we use square boxes to
represent latent variables and circles to represent observed variables.
\textup{(a)} DAG with latent variables; \textup{(b)} CPDAG; \textup{(c)}
PAG.}
\label{exampleintro}
\end{figure}

The first problem is that causal inference based on the PC algorithm
may be
incorrect. For example, consider the DAG in Figure
\ref{exampleintro}(a) with observed variables
$\mathbf{X}=\{X_1, X_2, X_3\}$ and latent variables $\mathbf{L}=\{L_1,
L_2\}$. The only conditional independence relationship among the
observed variables is $X_1 \ci X_3$. There
is only one DAG on $\mathbf{X}$ that implies this single conditional
independence relationship, namely $X_1 \rightarrow X_2 \leftarrow X_3$, and
this will therefore be the output of the PC algorithm; see Figure~\ref{exampleintro}(b).
This output would lead us to believe that both $X_1$ and $X_3$ are causes
of~$X_2$. But this is clearly incorrect, since in the underlying DAG with
latent variables, there is neither a~directed path from $X_1$ to $X_2$
nor one from $X_3$ to $X_2$.

%

A second problem is that the space of DAGs is not closed under
marginalization and conditioning~\cite{RichardsonSpirtes02} in the
following sense. If a~distribution is faithful to a~DAG, then the
distribution obtained by marginalizing out and conditioning on some of the
variables may \textit{not} be faithful to any DAG on the observed
variables. For example, consider the DAG $X_1 \rightarrow X_2 \leftarrow
L_1 \rightarrow X_3 \leftarrow X_4$. This DAG implies the following set
of conditional independence relationships among the observed variables
$\mathbf{X} = \{X_1,\ldots,X_4\}$:
$X_1 \ci X_3$, $X_1 \ci X_4$, $X_2 \ci X_4$, $X_1\ci X_3|X_4$, $X_1\ci
X_4|X_2$, $X_1\ci X_4|X_3$ and $X_2 \ci X_4|X_1$, and others implied by
these. There is no DAG on $\mathbf{X}$ that entails exactly this set of
conditional independencies via $d$-separation.

These problems can be solved by introducing a~new class of
graphs on the observed variables, called maximal ancestral graphs (MAGs)
\cite{RichardsonSpirtes02}. Every DAG with latent and
selection variables can be transformed into a~unique MAG
over the observed variables (\cite{RichardsonSpirtes02}, page 981).
Several DAGs can lead to the same MAG. In fact, a~MAG
describes infinitely many DAGs since no restrictions are made on the
number of
latent and selection variables.

MAGs encode causal relationships between the observed variables via the edge
marks. For example, consider the edge $X_1 \rightarrow
X_2$ in a~MAG. The tail at~$X_1$ implies that $X_1$ is a~cause (ancestor)
of $X_2$ or of a~selection variable, and the arrowhead at $X_2$ implies
that $X_2$ is not a~cause (not an ancestor) of~$X_1$ nor of any selection
variable, in all possible underlying DAGs with latent and selection
variables. Moreover, MAGs encode conditional independence relationships
among the observed variables via $m$-separation~\cite{RichardsonSpirtes02},
a~generalization of $d$-separation (see Definition~\ref{d-sepdefi} in
Section~\ref{secprobinterpretation}). Several MAGs can describe exactly the
same conditional independence relationships; see~\cite{Ali09equiv}. Such
MAGs form a~Markov equivalence class which can be represented by a~partial
ancestral graph (PAG); see Definition~\ref{defpagfci}.
PAGs describe causal features common to every MAG in the Markov
equivalence class,
and hence to every DAG (possibly with latent and selection variables)
compatible with the observable independence structure under the assumption
of faithfulness. For example, consider again the DAG in Figure
\ref{exampleintro}(a). The only conditional
independence relationship among the observed variables is $X_1 \ci X_3$,
and this is represented by the PAG in Figure
\ref{exampleintro}(c). This PAG
implies that $X_2$ is not a~cause (ancestor) of $X_1$, $X_3$ or a~selection
variable, and this is indeed the case in the underlying DAG in Figure
\ref{exampleintro}(a) and is true of any DAG that,
assuming faithfulness, could have implied $X_1 \ci X_3$.
The two circle marks at $X_1$ and $X_3$ in Figure
\ref{exampleintro}(c) represent uncertainty about
whether or not $X_1$ and $X_3$ are causes of $X_2$. This reflects the fact
that the single conditional independence relationship $X_1 \ci X_3$ among
the observed variables can arise from the DAG $X_1 \rightarrow X_2
\leftarrow X_3$ in which $X_1$ and $X_3$ are causes of $X_2$, but it
can also arise from the DAG in Figure
\ref{exampleintro}(a) in which $X_1$ and $X_3$ are
not causes of $X_2$.

Under the faithfulness assumption, a~Markov equivalence class of DAGs with
latent and selection variables can be learned from conditional
independence information among the observed variables alone using the Fast
Causal Inference (FCI) algorithm~\cite{SpirtesEtAl00}, which is a~modification of
the PC algorithm. Originally, the output of FCI was defined as a~partially oriented inducing path graph
(POIPG), but its output can also be interpreted as a~PAG
\cite{Zhang08-causal-reasoning-ancestral-graphs}. Spirtes et al.~\cite{SpirtesEtAl00} proved that the FCI
algorithm is sound in the presence of arbitrarily many latent
variables. Spirtes et al.~\cite{SpirtesMeekRichardson99} extended the soundness
proof to allow for selection variables as well. Zhang~\cite{Zhang08-orientation-rules} recently
introduced extra orientation rules that make FCI complete when its output
is interpreted as a~PAG. Despite its name, FCI is computationally very
intensive for large graphs.

Spirtes~\cite{Spirtes01-anytime} introduced a~modified version of FCI, called
Anytime FCI, that only considers conditional independence tests with
conditioning sets of size less
than some prespecified cut-off $K$. Anytime FCI is typically faster but
less informative than FCI, but the causal interpretation of tails and
arrowheads in its output is still sound.

Some work on the estimation of the size of causal effects in situations
with latent and selection variables can be found
in~\cite{RichardsonSpirtes03,Zhang08-causal-reasoning-ancestral-graphs} and in
Chapter~7 of~\cite{SpirtesEtAl00}.

\subsection{New contributions}\label{secintrocontributions}

We introduce a~new algorithm for learning PAGs, called the
\textit{Really} Fast Causal Inference (RFCI) algorithm (see Section
\ref{secrfci}). RFCI uses fewer conditional independence tests than FCI,
and its tests condition on a~smaller number of variables. As a~result, RFCI
is much faster than FCI and its output tends to be more reliable
for small samples, since conditional independence tests of high order
have low power. On the other hand, the output of RFCI may be less
informative. In this sense, the algorithm is related to the Anytime
FCI algorithm~\cite{Spirtes01-anytime}.

In Section~\ref{secrfci=fci} we compare the outputs of FCI and
RFCI, and define a~class of graphs for which the outputs of FCI and
RFCI are identical.

In Section~\ref{sechighdimconsistency} we prove consistency of FCI and
RFCI in sparse high-dimensional settings. The sparsity
conditions needed for consistency of FCI are stronger than those for RFCI,
due to the higher complexity of the FCI algorithm.

In order to compare RFCI to existing algorithms, we propose several
small modifications
of FCI and Anytime FCI. In particular, we introduce the Adaptive Anytime
FCI (AAFCI) algorithm (see Section 3 of the
supplementary document~\cite{ColomboEtAl12-supp}) and we propose several
ways to speed up the FCI and AAFCI algorithms (see Section~\ref{secfci}).

We show in simulations (see Section~\ref{secsimulations}) that the
numbers of errors made by all algorithms are very
similar. Moreover, we show that our modifications of FCI and AAFCI shorten
the computation time considerably, but that for large graphs, RFCI is
the only feasible algorithm.

All proofs, a~description of AAFCI, pseudocodes and two additional examples
are given in the supplementary document~\cite{ColomboEtAl12-supp}. The R-package
\texttt{pcalg}~\cite{KalischEtAl11} contains implementations of all
algorithms.

\section{Preliminaries}\label{secpreliminaries}

This section introduces terminology that is used throughout the
paper. Section~\ref{secgraphdef} defines various graphical concepts, and
Section~\ref{secprobinterpretation} describes how graphs can be
interpreted probabilistically and causally.

\subsection{Graphical definitions}\label{secgraphdef}

A graph $\mathcal{G}=(\mathbf{V},\mathbf{E})$ is composed of a~set of vertices
$\mathbf{V}=\{X_1,\ldots,X_p\}$ and a~set of edges $\mathbf{E}$.
In our framework, the vertices represent random variables and the edges
describe conditional independence and ancestral relationships.
The edge set $\mathbf{E}$ can contain (a~subset of) the following six types
of edges: $\rightarrow$ (\textit{directed}), $\leftrightarrow$ (\textit
{bi-directed}), $-$
(\textit{undirected}), $\twocirc$ (\textit{nondirected}), $\leftcirc$
(\textit{partially undirected}) and $\leftcircarrow$
(\textit{partially
directed}). The endpoints of an edge are called \textit{marks} and they
can be \textit{tails}, \textit{arrowheads} or \textit{circles}. We use the
symbol ``$*$'' to denote an arbitrary edge mark. A graph containing only
directed edges is called \textit{directed}, and one containing only
undirected edges is called
\textit{undirected}. A \textit{mixed} graph can contain directed,
bi-directed and undirected edges. If we are only interested in the
presence and absence of edges in a~graph and not in the edge marks, we
refer to the
\textit{skeleton} of the graph.

All the graphs we consider are \textit{simple} in that there is at most one
edge between any two vertices. If an edge is present, the vertices are said
to be \textit{adjacent}. If all pairs of vertices in a~graph are adjacent,
the graph is called \textit{complete}. The \textit{adjacency set} of a~vertex
$X_i$ in a~graph $\mathcal{G}$ is the set of all vertices in $\mathbf
{V}\setminus\{X_i\}$ that are
adjacent to $X_i$ in $\mathcal{G}$, denoted by adj$(\mathcal{G},X_i)$.
A~vertex $X_j$ in adj$(\mathcal{G},X_i)$ is called
a~\textit{parent} of $X_i$ if $X_j \rightarrow X_i$, a~\textit{child} of
$X_i$ if $X_i \rightarrow X_j$, a~\textit{spouse}
of $X_i$ if $X_i \leftrightarrow X_j$, and a~\textit{neighbor} of $X_i$ if
$X_i-X_j$. The corresponding sets of parents, children, spouses and
neighbors are denoted by pa$(\mathcal{G},X_i)$, ch$(\mathcal{G},X_i)$,
sp$(\mathcal{G},X_i)$ and ne$(\mathcal{G},X_i)$, respectively.

A \textit{path} is a~sequence of distinct adjacent
vertices. A path $\langle X_i, X_j, \ldots, X_k\rangle$ is said to be
\textit{out of} (\textit{into}) $X_i$ if the edge between $X_i$ and $X_j$ has a~tail
(arrowhead) at $X_i$. A \textit{directed path} is a~path along directed
edges that
follows the direction of the arrowheads. A \textit{cycle} occurs when there
is a~path from~$X_i$ to~$X_j$ and~$X_i$ and~$X_j$ are adjacent.
A directed path from $X_i$ to $X_j$ forms a~\textit{directed cycle}
together with the edge $X_j
\rightarrow X_i$, and it forms an \textit{almost directed cycle} together
with the edge $X_j \leftrightarrow
X_i$. If there is a~directed path $\pi$ from $X_i$ to $X_j$ or if
$X_i=X_j$, the vertex $X_i$ is called an
\textit{ancestor} of~$X_j$ and~$X_j$ a~\textit{descendant} of $X_i$. The
sets of ancestors and descendants of a~vertex $X_i$ in $\mathcal{G}$
are denoted
by an$(\mathcal{G},X_i)$ and de$(\mathcal{G},X_i)$, respectively. These
definitions are
applied to a~set $\mathbf{Y} \subseteq\mathbf{V}$ of distinct vertices
as follows:
\begin{eqnarray*}
\operatorname{an}(\mathcal{G},\mathbf{Y}) &=& \{X_i \vert X_i \in
\operatorname{an} (\mathcal{G},X_j)
\mbox{ for some } X_j \in\mathbf{Y}\};\\
\operatorname{de}(\mathcal{G},\mathbf{Y}) &=& \{X_i \vert X_i \in
\operatorname{de} (\mathcal{G},X_j)
\mbox{ for some } X_j \in\mathbf{Y}\}.
\end{eqnarray*}

Three vertices that form a~cycle are called a~\textit{triangle}. Three
vertices $\langle X_i,X_j, X_k\rangle$ are called an \textit{unshielded triple}
if $X_i$ and $X_j$ are adjacent, $X_j$ and $X_k$ are adjacent, but $X_i$
and $X_k$ are not adjacent. A nonendpoint vertex~$X_j$ on a~path $\pi$
is a~\textit{collider} on the
path if both the edges preceding and succeeding it have an arrowhead at
$X_j$, that is, if the path contains $\leftstar X_j \rightstar$. A
nonendpoint vertex $X_j$ on a~path $\pi$ which is not a~collider is a~\textit{noncollider} on the path. An unshielded
triple $\langle X_i,X_j,X_k\rangle$ is called a~\textit{v-structure}
if $X_j$
is a~collider on the path $\langle X_i,X_j,X_k\rangle$.\vadjust{\goodbreak}

A path $\pi=\langle X_l,\ldots,X_j,X_b,X_p\rangle$ in a~mixed graph is called
a~\textit{discriminating path for $X_b$} if the following three conditions
hold: (i) $\pi$ includes at least three edges;
(ii) $X_b$ is a~nonendpoint vertex on $\pi$ and is adjacent to $X_p$ on
$\pi$;
and (iii) $X_l$ is not adjacent to $X_p$ in the graph and every vertex
between~$X_l$ and~$X_b$ is a~collider on $\pi$ and a~parent of $X_p$. An example of a~discriminating path is given in Figure 4 of
\cite{ColomboEtAl12-supp}, where the circle marks are replaced by stars.

A graph $\mathcal{G}=(\mathbf{V},\mathbf{E})$ is called \textit
{connected} if
there exists a~path between any pair of vertices in $\mathbf{V}$.
A graph is called \textit{biconnected} if it is connected and remains so if
any vertex and its incident edges were to be removed. A~\textit{biconnected
component} of a~graph is a~maximally biconnected subgraph~\cite{Aho74}.

A directed graph $\mathcal{G}=(\mathbf{V},\mathbf{E})$ is called a~\textit{directed acyclic graph} (DAG) if
it does not contain directed cycles. A mixed graph
$\mathcal{G}=(\mathbf{V},\mathbf{E})$ is called an \textit{ancestral
graph} if
(i) it does not contain directed cycles, (ii) it does not contain
almost directed
cycles, and (iii) for any undirected edge $X_i-X_j$ in~$\mathbf{E}$,
$X_i$ and $X_j$ have
no parents or spouses. DAGs form a~subset of ancestral graphs.

\subsection{Probabilistic and causal interpretation of graphs}
\label{secprobinterpretation}

A DAG entails conditional independence relationships via a~graphical
criterion called $d$-se\-paration, which is a~special case of $m$-separation:
%
\begin{defi}[(Richardson and Spirtes~\cite{RichardsonSpirtes02})]\label{d-sepdefi}
 A path $\pi$ in an
ancestral graph is said to be blocked by a~set of vertices $\mathbf{Y}$
if and only if:
\begin{longlist}
\item $\pi$ contains a~subpath $\langle X_i, X_j, X_k\rangle$
such that the middle vertex $X_j$
is a~noncollider on this path and $X_j \in\mathbf{Y}$, or
\item $\pi$ contains a~v-structure $X_i \leftstar X_j
\rightstar
X_k$ such that $X_j \notin\mathbf{Y}$
and no descendant of $X_j$ is in $\mathbf{Y}$.
\end{longlist}
Vertices $Z$ and $W$ are $m$-separated
by $\mathbf{Y}$ if every path $\pi$ between $Z$
and $W$ is blocked by $\mathbf{Y}$. Sets of vertices $\mathbf{Z}$ and
$\mathbf{W}$ are $m$-separated by $\mathbf{Y}$ if all pairs of vertices $Z
\in\mathbf{Z}$, $W \in\mathbf{W}$ are $m$-separated by $\mathbf{Y}$.
\end{defi}

If two vertices $X_i$ and $X_j$ in a~DAG $\mathcal{G}$ are $d$-separated
by a~subset $\mathbf{Y}$ of the remaining vertices, then $X_i \ci X_j |
\mathbf{Y}$ in any distribution $Q$ that factorizes according to
$\mathcal{G}$ (i.e., the joint density can be written as the product of
the conditional
densities of each variable given its parents in $\mathcal G$:
$q(X_1,\ldots,X_p) = \prod_{i=1}^{p} q(X_i\vert{\operatorname
{pa}}(\mathcal{G}, X_i)$).
A~distribution $Q$ is said to be
\textit{faithful} to a~DAG $\mathcal{G}$ if the reverse implication also
holds, that is, if the conditional independence relationships in $Q$ are
exactly the same as those that can be inferred from~$\mathcal{G}$ using
$d$-separation. A set $\mathbf{Y}$ that $d$-separates $X_i$ and $X_j$ in a~DAG
is called a~\textit{minimal separating set} if no subset of $\mathbf{Y}$
$d$-separates $X_i$ and $X_j$. A~set~$\mathbf{Y}$ is a~\textit{minimal
separating set for $X_i$ and $X_j$ given $\mathbf{S}$} if $X_i$ and $X_j$
are $d$-separated by $\mathbf{Y}\cup\mathbf{S}$ and there is no subset
$\mathbf{Y}'$ of $\mathbf{Y}$ such that $X_i$ and $X_j$ are $d$-separated
by $\mathbf{Y}'\cup\mathbf{S}$.\vadjust{\goodbreak}

When a~DAG $\mathcal{G}=(\mathbf{V},\mathbf{E})$ contains latent and
selection variables, we write
$\mathbf{V}=\mathbf{X}\ddcup  \mathbf{L}\ddcup  \mathbf{S}$, where
$\mathbf{X}$ represents the observed variables, $\mathbf{L}$ represents the
latent variables and $\mathbf{S}$ represents the selection variables, and
these sets are disjoint (i.e., $\dot{\cup}$ denotes the union of disjoint
sets).

A \textit{maximal ancestral graph} (MAG) is
an ancestral graph in which every missing edge corresponds to a~conditional
independence relationship. Richardson and Spirtes (\cite
{RichardsonSpirtes02}, page 981) give an
algorithm to transform a~DAG $\mathcal{G}=(\mathbf{X}\ddcup
\mathbf{L}\ddcup  \mathbf{S},\mathbf{E})$ into a~unique MAG $\mathcal{G}^*$ as follows. Let $\mathcal{G}^*$ have vertex set
$\mathbf{X}$. For any pair of vertices $X_i, X_j \in\mathbf{X}$ make them
adjacent in~$\mathcal{G}^*$ if and only if there is an inducing path (see
Definition~\ref{defindpath}) between $X_i$ and~$X_j$ in $\mathcal{G}$
relative to $\mathbf{X}$ given $\mathbf{S}$. Moreover, for each edge $X_i
\twostar X_j$ in $\mathcal{G}^*$ put an arrowhead at $X_i$ if $X_i\notin
\operatorname{an}(\mathcal{G}, \{X_j\}\cup\mathbf{S})$ and put a~tail
otherwise. The resulting MAG $\mathcal{G}^* = ({\mathbf X},{\mathbf E}^*)$
encodes the
conditional independence relationships holding in $\mathcal{G}$ among
the observed
variables ${\mathbf X}$ conditional on some value for the selection variables
${\mathbf S}={\mathbf s}$; thus if $X_i$ and $X_j$ are $m$-separated by ${\mathbf Y}$ in
$\mathcal{G}^*$, then~$X_i$ and~$X_j$ are $d$-separated by ${\mathbf Y} \cup
{\mathbf S}$ in $\mathcal{G}$ and hence $X_i \ci X_j |({\mathbf Y}
\cup\{\mathbf{S}=\mathbf{s}\})$ in any distribution $Q$ factorizing according to
$\mathcal{G}$. Perhaps more importantly, an ancestral graph preserves the
ancestral relationships encoded in the DAG.

Throughout the remainder of this paper, $\mathbf{S}$
refers to either the set of variables $\mathbf{S}$ or the event
$\mathbf{S}=\mathbf{s}$, depending on the context.

\section{Oracle versions of the algorithms}\label{secalgorithms}

We consider the following problem: assuming that the distribution of
$\mathbf{V}= \mathbf{X}\ddcup  \mathbf{L}\ddcup  \mathbf{S}$ is
faithful to an
unknown underlying causal DAG $\mathcal{G}=(\mathbf{V},\mathbf{E})$,
and given oracle
information about all conditional independence relationships
between pairs of variables $X_i$ and $X_j$ in $\mathbf{X}$ given sets
$\mathbf{Y}\cup\mathbf{S}$ where $\mathbf{Y}\subseteq\mathbf
{X}\setminus\{X_i,X_j\}$,
we want to infer information about the ancestral (causal) relationships
of the variables in the
underlying DAG, which we represent via a~PAG.

We discuss and compare two algorithms for this purpose, the FCI
algorithm and our new RFCI algorithm. We first define the outputs of both
algorithms: an FCI-PAG and an RFCI-PAG. (An FCI-PAG is usually
referred to simply as a~``PAG,'' but in the remainder of this paper we use
the name FCI-PAG to make a~clear distinction between the output of the
two algorithms.)
%
\begin{defi}\label{defpagfci}
Let $\mathcal{G}$ be a~DAG with partitioned vertex set ${\mathbf X}\ddcup
{\mathbf L} \ddcup  {\mathbf S}$. Let $\mathcal C$ be a~simple
graph with vertex set ${\mathbf X}$ and edges of the type $\rightarrow$,
$\leftcircarrow$, $\twocirc$, $\leftrightarrow$, $-$ or $\leftcirc$.
Then $\mathcal C$ is said to be an FCI-PAG that represents $\mathcal
G$ if and only if, for any distribution $P$ of ${\mathbf X}\ddcup
{\mathbf L} \ddcup {\mathbf S}$ that is faithful to $\mathcal G$, the following
four conditions hold:
\begin{longlist}[(iii)]
\item[(i)] the absence of an edge between two vertices $X_i$ and $X_j$ in
$\mathcal{C}$ implies that there exists a~subset $\mathbf{Y} \subseteq
\mathbf{X}\setminus\{X_i,X_j\}$ such that $X_i \ci X_j |
(\mathbf{Y}\cup\mathbf{S})$ in~$P$;
\item[(ii)] the presence of an edge between two vertices $X_i$ and
$X_j$ in $\mathcal{C}$ implies that $X_i\ \newnci\ X_j | (\mathbf{Y}\cup
\mathbf{S})$ in $P$ for all subsets $\mathbf{Y}\subseteq\mathbf
{X}\setminus\{X_i,X_j\}$;
\item[(iii)] if an edge between $X_i$ and $X_j$ in $\mathcal{C}$ has
an arrowhead at $X_j$, then $X_j \notin\operatorname{an}(\mathcal
{G},X_i \cup\mathbf{S})$;
\item[(iv)] if an edge between $X_i$ and $X_j$ in $\mathcal{C}$ has a~tail at $X_j$, then $X_j \in\operatorname{an}(\mathcal{G},X_i \cup
\mathbf{S})$.
\end{longlist}
\end{defi}
%
\begin{defi}\label{defpagrfci}
Let $\mathcal{G}$ be a~DAG with partitioned vertex set ${\mathbf X}\ddcup
{\mathbf L} \ddcup {\mathbf S}$. Let $\mathcal C$ be a~simple
graph with vertex set ${\mathbf X}$ and edges of the type $\rightarrow$,
$\leftcircarrow$, $\twocirc$, $\leftrightarrow$, $-$, or $\leftcirc$.
Then $\mathcal C$ is said to be an RFCI-PAG that represents $\mathcal
G$ if and only if, for any distribution $P$ of ${\mathbf X}\ddcup
{\mathbf L} \ddcup {\mathbf S}$ that is faithful to $\mathcal G$, conditions~(i),
(iii) and (iv) of Definition
\ref{defpagfci} and the following condition hold:
\begin{longlist}[(ii$'$)]
\item[(ii$'$)] the presence of an edge between two vertices $X_i$ and
$X_j$ in $\mathcal{C}$ implies that $X_i\ \newnci\ X_j | (\mathbf{Y}\cup
\mathbf{S})$ for all subsets $\mathbf{Y}
\subseteq\operatorname{adj}(\mathcal{C},X_i)\setminus\{X_j\}$ and for all
subsets $\mathbf{Y}\subseteq\operatorname{adj}(\mathcal{C},X_j)\setminus
\{X_i\}$.
\end{longlist}
\end{defi}

Condition (ii) in Definition~\ref{defpagfci} is stronger than condition
(ii$'$) in Definition~\ref{defpagrfci}. Hence, the presence of an edge in an
RFCI-PAG has a~weaker interpretation than in an FCI-PAG. This has several
consequences. First, every FCI-PAG is an RFCI-PAG. Second, different
RFCI-PAGs for the same underlying DAG may have different skeletons, while
the FCI-PAG skeleton is unique. In general, the RFCI-PAG skeleton is a~supergraph of the FCI-PAG skeleton. Finally, an RFCI-PAG can correspond to
more than one Markov equivalence class of DAGs (see Example
\ref{exRFCInotFCI} in Section~\ref{secexample}).

It is worth noting that every FCI-PAG is an RFCI-PAG. Moreover, for a~given pair of a~graph $\mathcal{G}$ and a~distribution $P$ faithful to it,
there may be two different FCI-PAGs that represent $\mathcal{G}$ but
they will have the same
skeleton. On the other hand, for a~given pair of a~graph~$\mathcal{G}$
and a~distribution~$P$
faithful to it, there may also be more than one RFCI-PAG that represents~$\mathcal{G}$
and these different RFCI-PAGs can also have different skeletons.

The remainder of this section is organized as follows.
Section~\ref{secfci} briefly discusses the FCI algorithm and proposes
modifications that can speed up the algorithm while remaining sound and
complete.
Section~\ref{secrfci} introduces our new RFCI algorithm. Section
\ref{secexample} discusses several examples that illustrate the
commonalities and differences between the two algorithms, and Section
\ref{secrfci=fci} defines a~class of graphs for which the outputs of FCI
and RFCI are identical.

\subsection{The FCI algorithm}\label{secfci}

A high-level sketch of FCI (\cite{SpirtesEtAl00}, pages 144 and~145) is
given in
Algorithm~\ref{pseudofci}. The sub-algorithms 4.1--4.3 are given in~\cite{ColomboEtAl12-supp}.

\begin{algorithm}
\caption{The FCI algorithm}
\label{pseudofci}
{\fontsize{9pt}{11pt}\selectfont{
\begin{algorithmic}[1]
\REQUIRE Conditional independence information among all variables in
$\mathbf{X}$ given $\mathbf{S}$
\STATE Use Algorithm 4.1 of~\cite{ColomboEtAl12-supp} to
find an initial skeleton ($\mathcal{C}$), separation sets (sepset) and
unshielded triple list ($\mathfrak{M}$);
\STATE Use Algorithm 4.2 of
\cite{ColomboEtAl12-supp} to orient
v-structures (update $\mathcal{C}$);
\STATE Use Algorithm 4.3 of~\cite{ColomboEtAl12-supp}
to find the final
skeleton (update $\mathcal{C}$ and sepset);
\STATE Use Algorithm 4.2 of
\cite{ColomboEtAl12-supp} to orient
v-structures (update $\mathcal{C}$);
\STATE Use rules (R1)--(R10) of~\cite{Zhang08-orientation-rules} to
orient as many edge marks as possible (update $\mathcal{C}$);
\RETURN{$\mathcal{C}$, sepset}.
\end{algorithmic}}}
\end{algorithm}

The determination of adjacencies in the PAG within the FCI algorithm is
based on the following fact: if $X_i$ is not an ancestor of~$X_j$,
and~$X_i$ and~$X_j$ are conditionally\vadjust{\goodbreak}
independent given some set $\mathbf{Y}\cup\mathbf{S}$ where $\mathbf
{Y}\subseteq\mathbf{X}\setminus
\{X_i,X_j\}$, then $X_i$ and $X_j$ are conditionally independent given
$\mathbf{Y}^{\prime}\cup\mathbf{S}$ for some subset~$\mathbf{Y}^{\prime}$
of a~certain set $\mbox{D-SEP}(X_i,X_j)$ or of $\mbox{D-SEP}(X_j,X_i)$ (see
\cite{SpirtesEtAl00}, page~134 for a~definition). This means that, in
order to determine whether there is an
edge between $X_i$ and $X_j$ in an FCI-PAG, one does not need to test
whether $X_i \ci X_j | (\mathbf{Y} \cup\mathbf{S})$ for all possible
subsets $\mathbf{Y} \subseteq\mathbf{X} \setminus\{X_i,X_j\}$, but only
for all possible subsets $\mathbf{Y} \subseteq
\mbox{D-SEP}(X_i,X_j)$ and $\mathbf{Y} \subseteq
\mbox{D-SEP}(X_j,X_i)$. Since the sets D-SEP$(X_i,X_j)$ cannot be inferred
from the observed conditional independencies, Spirtes et al.
\cite{SpirtesEtAl00} defined a~superset, called Possible-D-SEP, that
can be computed:\vspace*{-2pt}
%
\begin{defi}\label{defpds}
Let $\mathcal{C}$ be a~graph with any of the following edge types:
$\twocirc$,
$\leftcircarrow$, $\leftrightarrow$. \textit
{Possible-D-SEP}$(X_i,X_j)$ in $\mathcal{C}$,
denoted in shorthand by $\operatorname{pds}(\mathcal{C},\break X_i,X_j)$, is defined as
follows: $X_k\in
\operatorname{pds}(\mathcal{C},X_i,X_j)$ if and only if there is
a~path $\pi$ between $X_i$ and $X_k$ in $\mathcal{C}$ such that for every
subpath $\langle X_m,X_l,X_h\rangle$ of $\pi$, $X_l$ is a~collider on
the subpath in
$\mathcal{C}$ or $\langle X_m, X_l,X_h\rangle$ is a~triangle in
$\mathcal{C}$.\vspace*{-2pt}
\end{defi}

\begin{remark}
Note that $X_j$ does not play a~role in the definition of
$\operatorname{pds}(\mathcal{C}$, $X_i,X_j)$, but we keep it as an argument
because we will later consider alternative definitions of Possible-D-SEP
(see Definition~\ref{defpdspath}) where the second vertex $X_j$ does
play a~role.\vspace*{-2pt}
\end{remark}

Since the definition of Possible-D-SEP requires some knowledge about the
skeleton and orientation of edges, the FCI algorithm first finds an initial
skeleton denoted by $\mathcal{C}_1$ in Step 1. This is done as in the
PC-algorithm, by starting with a~complete graph with edges $\twocirc$
and performing conditional independence tests given subsets of
increasing size of the adjacency sets of the vertices. An edge between
$X_i$ and $X_j$ is deleted if a~conditional independence is found, and
the set responsible for this conditional independence is saved in
$\operatorname{sepset}(X_i,X_j)$ and $\operatorname{sepset}(X_j,X_i)$ (see Algorithm
4.1 of~\cite{ColomboEtAl12-supp}). The skeleton
after completion of Step 1 is a~superset of the final skeleton.

In Step 2, the algorithm orients unshielded triples $X_i
\leftstarrightcirc X_j \rightstarleftcirc X_k$ as v-structures $X_i
\leftstar X_j \rightstar X_k$ if and only if $X_j$ is not in
$\operatorname{sepset}(X_i,X_k)$ and
$\operatorname{sepset}(X_k,X_i)$ (see Algorithm 4.2 of~\cite{ColomboEtAl12-supp}).

The graph resulting after Step 2, denoted by $\mathcal{C}_2$, contains
sufficient information\vadjust{\goodbreak} to compute the Possible-D-SEP sets. Thus, in
Step 3, the algorithm computes $\operatorname{pds}(\mathcal{C}_2,X_i,\cdot)$ for every
$X_i \in
\mathbf{X}$. Then for every element $X_j$ in adj$(\mathcal{C}_2,X_i)$, the
algorithm tests whether $X_i \ci X_j | (\mathbf{Y} \cup\mathbf{S})$ for
every subset~$\mathbf{Y}$ of
$\operatorname{pds}(\mathcal{C}_2,X_i,\cdot)\setminus\{X_i,X_j\}$ and of $\operatorname{pds}(\mathcal
{C}_2,X_j,\cdot)\setminus\{X_j,X_i\}$
(see Algorithm~4.3 of~\cite{ColomboEtAl12-supp}).
As in Step 1,
the tests are arranged in a~hierarchical way
starting with conditioning sets of small size. If there exists a~set
$\mathbf{Y}$
that makes $X_i$ and $X_j$ conditionally independent given $\mathbf{Y}
\cup\mathbf{S}$, the edge between~$X_i$ and~$X_j$ is
removed and the set $\mathbf{Y}$ is saved as the separation set in
$\operatorname{sepset}(X_i,X_j)$ and $\operatorname{sepset}(X_j,X_i)$. After all conditional
independence tests are
completed, every edge in $\mathcal{C}$ is reoriented as $\twocirc$,
since the
orientation of v-structures in Step 2 of the algorithm
cannot necessarily be interpreted as specified in conditions (iii) and (iv)
of Definition~\ref{defpagfci}.

In Step 4, the v-structures are therefore oriented again based on the
updated skeleton and the updated information
in sepset (see Algorithm 4.2 of~\cite{ColomboEtAl12-supp}). Finally, in Step 5
the algorithm replaces as many circles as possible by arrowheads and
tails using the orientation
rules described by~\cite{Zhang08-orientation-rules}.

\subsubsection*{First proposed modification: FCI$_{\mathrm{path}}$}
For sparse graphs, Step 3 of the FCI algorithm dramatically increases
the computational complexity of the algorithm when compared
to the PC algorithm. The additional computational effort can be
divided in two parts: computing the Possible-D-SEP sets, and testing conditional
independence given all subsets of these sets.
The latter part is computationally infeasible when the sets
$\operatorname{pds}(\mathcal{C}_2,X_i,\cdot)$ are large,
containing, say, more than 30 vertices. Since the size of the
Possible-D-SEP sets plays such an important role in the complexity of the
FCI algorithm, and since one has some freedom in defining these sets (they
simply must be supersets of the D-SEP sets),
we first propose a~modification of the definition of Possible-D-SEP
that can
decrease its size.
%
\begin{defi}\label{defpdspath}
Let $\mathcal{C}$ be a~graph with any of the following edge types:
$\twocirc$,
$\leftcircarrow$, $\leftrightarrow$. Then, for two vertices $X_i$ and
$X_j$ adjacent in $\mathcal{C}$,
$\mathrm{pds}_{\mathrm{path}}(\mathcal{C},X_i,X_j)$ is defined as
follows: $X_k\in
\mathrm{pds}_{\mathrm{path}}(\mathcal{C},X_i,  X_j)$ if and
only if (i) there is
a~path~$\pi$ between~$X_i$ and~$X_k$ in $\mathcal{C}$ such that for every
subpath $\langle X_m,X_l,X_h\rangle$ of $\pi$, $X_l$ is a~collider on the
subpath in $\mathcal{C}$ or
$\langle X_m, X_l,X_h\rangle$ is a~triangle in $\mathcal{C}$, and
(ii)~$X_k$ lies on a~path between $X_i$ and $X_j$.
\end{defi}

For any pair of adjacent vertices $X_i$ and $X_j$ in a~graph $\mathcal C$,
the set $\mathrm{pds}_{\mathrm{path}} (\mathcal{C},\allowbreak X_i$, $X_j)$
can be computed easily by intersecting $\operatorname{pds}(\mathcal{C},
X_i,\cdot)$ with
the unique biconnected component in $C$ that contains the edge between
$X_i$ and $X_j$.
Algorithm~4.3 of~\cite{ColomboEtAl12-supp} can now
be modified as follows. Before line 1,
we compute all biconnected components of the graph $\mathcal{C}_2$,
where $\mathcal C_2$ is the graph resulting from Step 2 of the FCI algorithm.
Then between lines 3 and 4, we compute $\mathrm{pds}_{\mathrm
{path}}(\mathcal{C}_2,X_i,X_j)$ as described above.
Finally, on lines 8, 13 and 14, we replace
$\mathrm{pds}(\mathcal{C}_2,X_i,\cdot)$ by
$\mathrm{pds}_{\mathrm{path}}(\mathcal{C}_2,X_i,X_j)$. We refer to the FCI
algorithm with this modified version of Algorithm
4.3 of~\cite{ColomboEtAl12-supp} as FCI$_{\mathrm{path}}$.

\subsubsection*{Second class of modifications: CFCI, CFCI$_{\mathrm
{path}}$, SCFCI
and SCFCI$_{\mathrm{path}}$}
Another possibility to decrease the size of Possible-D-SEP is to use
conservative rules to
orient v-structures in Step 2 of the FCI algorithm, so that fewer
arrowheads are introduced, similarly to the
Conservative PC algorithm~\cite{RamseyZhangSpirtes06}. This is
especially helpful in the sample version of the algorithm (see
Section~\ref{secsampleversionalgorithms}), as the sample version tends to
orient too many v-structures, which can lead to long chains of
bi-directed edges
and hence large Possible-D-SEP sets (see Figure~\ref{figsizepds} in
Section~\ref{seccomputingtime}).

The conservative orientation works as
follows. For all unshielded triples $\langle X_i,X_j,X_k\rangle$ in
$\mathcal
C_1$, where $\mathcal C_1$ is the graph resulting from Step 1 of the FCI
algorithm, we determine all subsets $\mathbf{Y}$ of
$\operatorname{adj}(\mathcal{C}_1,X_i)$ and of $\operatorname
{adj}(\mathcal{C}_1,X_k)$ satisfying $X_i \ci
X_k | (\mathbf{Y} \cup\mathbf{S})$. We refer to these sets as separating
sets, and we label the triple $\langle X_i,X_j,X_k\rangle$ as \textit
{unambiguous} if
and only if (i) at least one separating set $\mathbf{Y}$ is found and
either $X_j$ is in all separating sets and in $\operatorname
{sepset}(X_i,X_k)$ or $X_j$ is in
none of the separating sets nor in $\operatorname{sepset}(X_i,X_k)$, or (ii)
no such separating set $\mathbf{Y}$ is found.
[Condition (ii) can occur, since separating sets found in Step 1 of the
FCI algorithm do not need to be a~subset of $\operatorname{adj}(\mathcal
{C}_1,X_i)$ or
of $\operatorname{adj}(\mathcal{C}_1,X_k)$.] At the end of Step 2, we
only orient
unambiguous triples satisfying $X_j \notin\operatorname
{sepset}(X_i,X_k)$ as
v-structures. This may lead to different Possible-D-SEP sets in Step
3 (even in the oracle version of the algorithm), but other than that,
Steps 3--5 of the algorithm remain unchanged.
We refer to this version of the FCI algorithm as Conservative FCI
(CFCI). If CFCI is used in combination with $\mathrm{pds}_{\mathrm{path}}$,
we use the name CFCI$_{\mathrm{path}}$.

Finally, the idea of conservative v-structures can also be applied in
Step 4 of the FCI algorithm. For each unshielded triple
$\langle X_i,X_j,X_k\rangle$ in $\mathcal C_3$, where $\mathcal{C}_3$
is the
graph resulting from Step 3, we determine all subsets $\mathbf{Y}$
of $\operatorname{adj}(\mathcal{C}_3,X_i)$ and of $\operatorname
{adj}(\mathcal{C}_3,X_k)$ satisfying $X_i
\ci X_k | (\mathbf{Y} \cup\mathbf{S})$. We then determine if a~triple
is unambiguous,
and only if this is the case we orient it as v-structure or
non-v-structure. Moreover, the orientation rules
in Step 5 of the algorithm are adapted so that they only rely on
unambiguous triples. We use the name
Superconservative FCI (SCFCI) to refer to the version of FCI that uses
conservative v-structures in both Steps 2 and 4. If SCFCI is used in
combination with
$\mathrm{pds}_{\mathrm{path}}$, we use the name SCFCI$_{\mathrm{path}}$.
The proof of Theorem~\ref{thsoundnessfciversions} shows that the
output of the oracle version of SCFCI is identical to that of CFCI. We
still consider this version, however, in the hope to obtain better edge
orientations in the sample versions of the algorithms, where
the outputs are typically not identical.

Soundness of FCI follows from Theorem 5 of~\cite{SpirtesMeekRichardson99}.
Soundness results for the modifications FCI$_{\mathrm{path}}$, CFCI,
CFCI$_{\mathrm{path}}$, SCFCI and SCFCI$_{\mathrm{path}}$
are given in the following theorem:
%
\begin{theo}\label{thsoundnessfciversions}
Consider one of the oracle versions of FCI$_{\mathrm{path}}$, CFCI,
CFCI$_{\mathrm{path}}$, SCFCI or SCFCI$_{\mathrm{path}}$. Let
the\vadjust{\goodbreak}
distribution of
$\mathbf{V} = \mathbf{X} \ddcup \mathbf{L}\ddcup \mathbf{S}$ be
faithful to a~DAG $\mathcal{G}$ and let conditional independence
information among all variables in $\mathbf{X}$ given
$\mathbf{S}$ be the input to the algorithm. Then the output of the
algorithm is an FCI-PAG of $\mathcal{G}$.
\end{theo}

Completeness of FCI was proved by~\cite{Zhang08-orientation-rules}.
This means that the output of FCI is maximally informative, in the
sense that for every circle mark there exists at least one MAG in the
Markov equivalence
class represented by the PAG where the mark is oriented as a~tail, and
at least one where it
is oriented as an arrowhead. Completeness results of FCI$_{\mathrm{path}}$, CFCI,
CFCI$_{\mathrm{path}}$, SCFCI and SCFCI$_{\mathrm{path}}$ follow directly
from the fact that, in the oracle versions, the orientation rules of these
modifications boil down to the orientation rules of FCI.

\subsection{The RFCI algorithm}\label{secrfci}

The Really Fast Causal Inference (RFCI)
algorithm is a~modification of FCI. The main
difference is that RFCI avoids the conditional independence tests
given subsets of Possible-D-SEP sets, which can become very large even for
sparse graphs. Instead, RFCI performs some additional tests
before orienting v-structures and discriminating paths in order to
ensure soundness, based on Lemmas
\ref{corrvstruct} and~\ref{corrdiscrpaths} below. The number of these
additional tests and the size of their conditioning sets is small for
sparse graphs, since RFCI only conditions on subsets of the adjacency
sets. As a~result, RFCI is much faster than FCI for sparse graphs (see
Section~\ref{seccomputingtime}). Moreover, the lower computational
complexity of RFCI leads to high-dimensional consistency results under
weaker conditions than FCI [compare conditions (A3) and (A3$'$) in Sections
\ref{secconsistrfci} and~\ref{secconsistFCI}].
A high-level sketch of RFCI is given in Algorithm
\ref{pseudorfci}.

\begin{algorithm}[b]
\caption{The RFCI algorithm}
\label{pseudorfci}
{\fontsize{9pt}{11pt}\selectfont{
\begin{algorithmic}[1]
\REQUIRE Conditional independence information among all variables in
$\mathbf{X}$ given $\mathbf{S}$
\STATE Use Algorithm 4.1 of~\cite{ColomboEtAl12-supp} to
find an initial skeleton ($\mathcal{C}$), separation sets (sepset) and
unshielded triple list ($\mathfrak{M}$);
\STATE Use Algorithm 4.4 of~\cite{ColomboEtAl12-supp} to orient
v-structures (update $\mathcal{C}$ and sepset);
\STATE Use Algorithm 4.5 of~\cite{ColomboEtAl12-supp} to orient
as many edge marks as possible (update $\mathcal{C}$ and sepset);
\RETURN{$\mathcal{C}$, sepset}.
\end{algorithmic}}}
\end{algorithm}

Step 1 of the algorithm is identical to Step 1 of Algorithm
\ref{pseudofci}, and is used to find an initial skeleton $\mathcal
{C}_1$ that
satisfies conditions (i) and (ii$'$) of Definition~\ref{defpagrfci}.

In Step 2 of the algorithm, unshielded triples are oriented based on
Lem\-ma~\ref{corrvstruct} and some further edges may be removed.
%
\begin{lemma}[(Unshielded triple rule)]\label{corrvstruct}
$\!\!\!$Let the distribution of $\mathbf{V}=\mathbf{X}\ddcup \mathbf{L}\ddcup
\mathbf{S}$ be faithful to a~DAG $\mathcal{G}$.
Assume that \textup{(a1)} $\mathbf{S}_{ik}$ is a~minimal separating set for
$X_i$ and
$X_k$ given $\mathbf{S}$, and \textup{(a2)} $X_i$ and $X_j$ as well as $X_j$
and $X_k$ are
conditionally dependent given $(\mathbf{S}_{ik}\setminus\{X_j\})
\cup\mathbf{S}$. Then
$X_j \in\operatorname{an}(\mathcal{G}, \{X_i,X_k\}\cup\mathbf{S})$
if and only if
$X_j \in\mathbf{S}_{ik}$.
\end{lemma}

The details of Step 2 are given in Algorithm 4.4 of
\cite{ColomboEtAl12-supp}. We start with a~list $\mathfrak{M}$ of all
unshielded triples in $\mathcal C_1$, where $\mathcal C_1$ is the graph
resulting from Step~1 of the RFCI algorithm, and an empty list
$\mathfrak{L}$ that is used to store triples that were found to satisfy
the conditions of Lemma~\ref{corrvstruct}. For each triple $\langle
X_i,X_j,X_k\rangle$ in~$\mathfrak{M}$, we check if both $X_i$ and $X_j$
and $X_j$ and $X_k$ are conditionally dependent given
$(\operatorname{sepset}(X_i,X_k)\setminus \{X_j\})\cup\mathbf{S}$.
These conditional dependencies may not have been checked in Step 1 of
the algorithm, since $\operatorname {sepset}(X_i,X_k)\setminus \{X_j\}$
does not need to be a~subset of $\operatorname{adj}(\mathcal
{C}_1,X_j)$. If both conditional dependencies hold, the triple
satisfies the conditions of Lemma~\ref{corrvstruct} and is added to
$\mathfrak{L}$. On the other hand, an additional conditional
independ\-ence relationship may be detected, say \mbox{$X_i \ci X_j |
((\operatorname{sepset}(X_i,X_k)\setminus\{X_j\})\cup\mathbf{S})$}. This
may arise in a~situation where $X_i$ and $X_j$ are not $m$-separated
given a~subset of vertices adjacent to $X_i$, and are not $m$-separated
given a~subset of vertices adjacent to $X_j$, but they do happen to be
$m$-separated given the set $(\operatorname {sepset}(X_i,X_k)\setminus
\{X_j\})\cup\mathbf{S}$. In this situation, we remove the edge $X_i
\twostar X_j$ from the graph, in agreement with condition (i) of
Definition~\ref{defpagrfci}. The removal of this edge can create new
unshielded triples, which are added to $\mathfrak{M}$. Moreover, it can
destroy unshielded triples in $\mathfrak{L}$ and $\mathfrak{M}$, which
are therefore removed. Finally, by testing subsets of the conditioning
set which led to removal of the edge, we find a~\textit{minimal}
separating set for $X_i$ and~$X_j$ and store it in $\operatorname{sepset}(X_i,X_j)$
and $\operatorname{sepset}(X_j,X_i)$. Example 1 of~\cite{ColomboEtAl12-supp}
shows that it is not sufficient to simply
store $\operatorname{sepset}(X_i,X_k) \setminus\{X_j\}$ since it may
not be minimal for $X_i$ and $X_j$. We work with the lists
$\mathfrak{M}$ and $\mathfrak{L}$ to ensure that the result of Step 2
does not depend on the order in which the unshielded triples are
considered.

After Step 2, all unshielded triples still present in the graph are
correctly oriented as a~v-structure or non-v-structure.
In Step 3, the algorithm orients as many further edges as possible, as
described in Algorithm 4.5 of
\cite{ColomboEtAl12-supp}. This procedure consists of repeated
applications of the orientation rules
(R1)--(R10) of~\cite{Zhang08-orientation-rules}, with the difference
that rule (R4) about
the discriminating path has been modified according to Lemma
\ref{corrdiscrpaths}.
%
\begin{lemma}[(Discriminating path rule)]\label{corrdiscrpaths}
Let the distribution of $\mathbf{V}=\mathbf{X}\ddcup
\mathbf{L}\ddcup \mathbf{S}$ be faithful to a~DAG $\mathcal{G}$.
Let $\pi_{ik}=\langle X_i,\ldots,X_l,X_j,X_k\rangle$ be a~sequence of
at least four
vertices that satisfy:
\textup{(a1)} $X_i$ and $X_k$ are conditionally independent given
$\mathbf{S}_{ik} \cup
\mathbf{S}$, \textup{(a2)} any two successive vertices $X_h$ and $X_{h+1}$
on~$\pi_{ik}$
are conditionally dependent given $(\mathbf{Y}' \setminus
\{X_h,X_{h+1}\})\cup\mathbf{S}$ for all
$\mathbf{Y}'\subseteq\mathbf{S}_{ik}$, \textup{(a3)} all vertices $X_h$
between $X_i$ and
$X_j$ (not including $X_i$ and $X_j$) satisfy $X_h \in\operatorname
{an}(\mathcal{G},
X_k)$ and $X_h \notin\operatorname{an}(\mathcal{G},\{X_{h-1},X_{h+1}\}
\cup\mathbf{S})$,
where $X_{h-1}$ and $X_{h+1}$ denote the vertices adjacent to $X_h$ on
$\pi_{ik}$. Then the following hold:
\textup{(b1)} if $X_j \in\mathbf{S}_{ik}$, then $X_j \in\operatorname
{an}(\mathcal{G},
\{X_k\} \cup\mathbf{S})$ and
$X_k \notin\operatorname{an}(\mathcal{G},\{X_j\} \cup\mathbf{S})$,
and \textup{(b2)} if $X_j
\notin\mathbf{S}_{ik}$,
then $X_j \notin\operatorname{an}(\mathcal{G},\{X_l,X_k\} \cup\mathbf
{S})$ and
$X_k \notin\operatorname{an}(\mathcal{G},\{X_j\} \cup\mathbf{S})$.
\end{lemma}

Lemma~\ref{corrdiscrpaths} is applied as follows. For each triangle
$\langle X_l,X_j,X_k\rangle$ of the form $X_j
\rightstarleftcirc X_k$, $X_j \leftstar X_l$ and $X_l \rightarrow
X_k$, the algorithm searches for a~discriminating path $\pi=
\langle X_i,\ldots,X_l,X_j,X_k\rangle$ for $X_j$ of minimal length, and checks
that the vertices in every consecutive pair $(X_r,X_q)$ on $\pi$ are
conditionally dependent given
$\mathbf{Y} \cup\mathbf{S}$ \textit{for all
subsets} $\mathbf{Y}$ of $\operatorname{sepset}(X_i,X_k)\setminus\{X_r,X_q\}$.
(Example 2 of~\cite{ColomboEtAl12-supp} shows why it is not
sufficient to only check conditional dependence given
$(\operatorname{sepset}(X_i,X_k)\setminus\{X_r,X_q\}) \cup\mathbf{S}$,
as we did for the v-structures.)
If we do not find any conditional independence relationship, the path
satisfies the conditions of Lemma~\ref{corrdiscrpaths} and is
oriented as in rule (R4) of~\cite{Zhang08-orientation-rules}. If one
or more conditional independence relationships are found, the
corresponding edges are removed, their minimal separating sets are
stored, and any new unshielded triples that are created by removing
the edges are oriented using Algorithm 4.4 of
\cite{ColomboEtAl12-supp}. We
note that the output of Step 3 may depend on the order in which the
discriminating paths are considered.

Soundness of RFCI is stated in the following theorem.
%
\begin{theo}\label{corrrfcitheo}
Let the distribution of $\mathbf{V} = \mathbf{X} \ddcup \mathbf{L}
\ddcup \mathbf{S}$ be faithful to\break a~DAG~$\mathcal{G}$ and
let conditional independence information among all variables in $\mathbf
{X}$ given~$\mathbf{S}$ be the input to the RFCI algorithm. Then the output of
RFCI is an RFCI-PAG of $\mathcal{G}$.
\end{theo}
%
\begin{remark}
The new orientation rules based on Lemmas~\ref{corrvstruct} and
\ref{corrdiscrpaths} open possibilities for different modifications of
the FCI algorithm. For example, one could replace
$\operatorname{pds}(\mathcal{C},X_i,X_j)$ by $\mathrm{pds}_k(\mathcal{C},X_i,X_j)$,
where a~vertex~$X_l$ is in
$\mathrm{pds}_k(\mathcal{C},X_i,X_j)$ if it is in $\operatorname
{pds}(\mathcal{C},X_i,X_j)$ and there
is a~path between~$X_i$ and~$X_l$ containing no more than $k+1$
vertices. This modification yields a~skeleton that is typically a~superset
of the skeleton of the true FCI-PAG. In order to infer correct causal
orientations based on this skeleton, one needs to use Algorithms
4.4 and~4.5 of~\cite{ColomboEtAl12-supp}
to determine the final orientations of the edges. The parameter $k$
represents a~trade-off between computing time and informativeness of the output, where
$k=1$ corresponds to the RFCI algorithm and $k=|\mathbf{X}|-2$
corresponds to the FCI algorithm.

Another way to obtain a~more informative but slower version of RFCI can be
obtained by modifying Step 1 of the RFCI algorithm: instead of considering
all subsets of $\operatorname{adj}(\mathcal{C},X_i)$ and of
$\operatorname{adj}(\mathcal{C},X_j)$, one can
consider all subsets of the union $\operatorname{adj}(\mathcal{C},X_i)
\cup\operatorname{adj}(\mathcal{C},X_j)$.
\end{remark}

\subsection{Examples}\label{secexample}
We now illustrate the algorithms in two examples. In Example
\ref{exRFCI=FCI}, the outputs of FCI and RFCI are identical. In
Example~\ref{exRFCInotFCI}, the outputs of FCI and RFCI are not identical,
and the output of RFCI describes two Markov\vadjust{\goodbreak} equivalence classes. We
will see, however, that the
ancestral or causal information inferred from an RFCI-PAG is correct. Two
additional examples illustrating details of Algorithms
4.4 and 4.5 of~\cite{ColomboEtAl12-supp} are
given in Section 5 of
\cite{ColomboEtAl12-supp}.\vspace*{-2pt}

\begin{example}\label{exRFCI=FCI}
Consider the DAG in Figure~\ref{imprgraphs}(a)
containing observed variables $\mathbf{X}=\{X_1,\ldots,X_6\}$, latent
variables $\mathbf{L}=\{L_1,L_2\}$ and no selection variables
($\mathbf{S}=\varnothing$). Suppose that all conditional independence
relationships over~$\mathbf{X}$ that can be read off from this DAG are used
as input for the algorithms.

%
\begin{figure}
\begin{tabular}{@{}ccc@{}}

\includegraphics{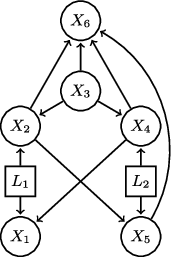}
 & \includegraphics{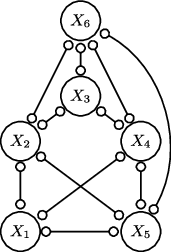} & \includegraphics{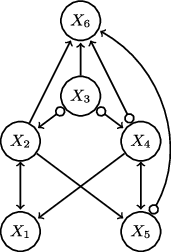}\\
(a) & (b) & (c)\vspace*{-3pt}
\end{tabular}
\caption{Graphs corresponding to Example \protect\ref{exRFCI=FCI},
where the
outputs of FCI and RFCI are identical. \textup{(a)}~Underlying DAG with
latent variables;
\textup{(b)} initial skeleton $\mathcal{C}_1$; \textup{(c)} RFCI-PAG
and FCI-PAG.}
\label{imprgraphs}
\vspace*{-2pt}
\end{figure}

In all algorithms, Step 1 is the same, and consists of finding an initial
skeleton. This skeleton, denoted by $\mathcal{C}_1$, is shown in Figure
\ref{imprgraphs}(b). The final output given by both
algorithms is shown in Figure~\ref{imprgraphs}(c).

Comparing the initial skeleton with the final skeleton, we see that the
edge $X_1
\twocirc X_5$ is present in the initial but not in the final skeleton. The
absence in the final skeleton is due to the fact that $X_1\ci X_5 |
\{X_2,X_3,X_4\}$. The edge is present in the initial skeleton, since
this conditional
independence is not found in Step~1 of the algorithms, because
$\{X_2,X_3,X_4\}$ is not a~subset of $\operatorname{adj}(\mathcal
{C}_1,X_1)$ nor of
$\operatorname{adj}(\mathcal{C}_1,X_5)$.

The FCI algorithm finds the conditional independence relationship $X_1
\ci
X_5 | \{X_2,X_3,X_4\}$
in Step 3 when subsets of Possible-D-SEP are considered, since
$\operatorname{pds}(\mathcal{C}_2,X_1,X_5)\setminus\{X_1,X_5\}=\{
X_2,X_3,X_4\}$ and
$\operatorname{pds}(\mathcal{C}_2,X_5,X_1)\setminus\{X_5, X_1\}=\{
X_2,X_3,X_4,X_6\}$, where
$\mathcal{C}_2$ is the graph resulting from Step 2 of the algorithm.

In the RFCI algorithm, the conditional independence relationship
$X_1\,{\ci}\,
X_5 |\allowbreak \{X_2$, $X_3,X_4\}$ is also found, but by another
mechanism. In Step 2 of Algorithm~\ref{pseudorfci},
unshielded triples are oriented after performing some additional
conditional independence tests. In particular, when considering the triple
$\langle X_1,X_5,X_6\rangle$, the algorithm checks whether $X_1\ci X_5 |
(\operatorname{sepset}(X_1,X_6) \setminus\{X_5\})$, where
$\operatorname{sepset}(X_1,X_6) = \{X_2,X_3,X_4\}$.\vadjust{\goodbreak}

This example also shows why it is necessary to check unshielded triples
according to Lemma~\ref{corrvstruct} before orienting them as
v-structures. Omitting this check for triple $\langle
X_1,X_5,X_6\rangle$ would orient
it as a~v-structure, since $X_5 \notin\operatorname{sepset}(X_1,X_6)$.
Hence, we
would conclude that $X_5 \notin\operatorname{an}(\mathcal{G},\{X_6\}
\cup
\mathbf{S})$, which contradicts the underlying DAG.

Finally, we see that the orientations of the edges are identical in the
outputs of both algorithms, which implies that the outputs encode the
same ancestral information.\vspace*{-3pt}
\end{example}
%
\begin{example}\label{exRFCInotFCI}
Consider the DAG $\mathcal{G}$ in Figure
\ref{graphs}(a), containing observed variables
$\mathbf{X}=\{X_1,\ldots,X_5\}$, latent variables
%
\begin{figure}
\begin{tabular}{@{}ccc@{}}

\includegraphics{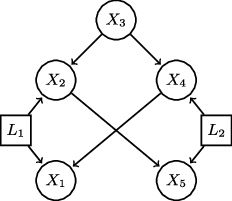}
 & \includegraphics{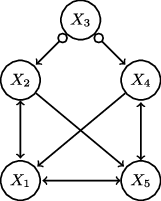} & \includegraphics{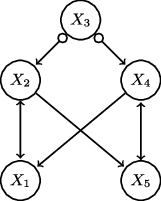}\\
(a) & (b) & (c)
\end{tabular}
\caption{Graphs corresponding to Example \protect\ref{exRFCInotFCI},
where the
outputs of FCI and RFCI are not identical. The output of RFCI corresponds
to two Markov equivalence classes when interpreted as an RFCI-PAG.
\textup{(a)} Underlying DAG $\mathcal{G}$ with latent variables;
\textup{(b)} output of RFCI for $\mathcal{G}$;
\textup{(c)} output of FCI for~$\mathcal{G}$.}
\label{graphs}
\end{figure}
$\mathbf{L}=\{L_1,L_2\}$ and no selection variables
($\mathbf{S}=\varnothing$) (see also
\cite{SpirtesMeekRichardson99}, page 228, Figure 8a). Suppose that all
conditional independence relationships over $\mathbf{X}$ that can be read
off from this DAG are used as input for the algorithms.

The outputs of the RFCI and FCI algorithms are shown in Figure
\ref{graphs}(b) and~(c),
respectively. We see that the output of RFCI contains an extra edge,
namely \mbox{$X_1 \leftrightarrow X_5$}.

As in Example~\ref{exRFCI=FCI}, this edge is present after Step 1 of both
algorithms. The reason is that the conditional independence $X_1 \ci
X_5 |
\{X_2,X_3,X_4\}$ is not found, because $\{X_2,X_3,X_4\}$ is not a~subset of
$\operatorname{adj}(\mathcal{C}_1,X_1)$ nor of $\operatorname
{adj}(\mathcal{C}_1,X_5)$,
where $\mathcal{C}_1$ denotes the skeleton after Step 1.

The FCI algorithm finds this conditional independence in Step 3 when
subsets of Possible-D-SEP are considered. The RFCI algorithm does not
find this conditional independence,
since the edge between $X_1$ and $X_5$ does not appear in an unshielded
triple or a~discriminating path. However, the ancestral information encoded
by the output of RFCI is correct, and in this example
identical to the ancestral information encoded by the output of
FCI.\looseness=-1

Finally, we show that the RFCI-PAG in Figure
\ref{graphs}(b) describes two Markov equivalence
classes. Consider a~new DAG $\mathcal{G}'$, which is adapted from
$\mathcal{G}$\vadjust{\goodbreak} in Figure
\ref{graphs}(a) by adding one additional latent variable $L_3$
pointing at $X_1$ and~$X_5$. This modification implies that $X_1$ and $X_5$
are conditionally dependent given any subset of the remaining observed
variables, so that $\mathcal{G}'$ belongs to a~different Markov
equivalence class than
$\mathcal{G}$. The output of both FCI and RFCI, when using as input the
conditional independence relationships that can be read off from
$\mathcal{G}'$, is given in Figure~\ref{graphs}(b). Hence,
the PAG in Figure~\ref{graphs}(b) represents more than one
Markov equivalence class if interpreted as an RFCI-PAG.
\end{example}

\subsection{A class of graphs for which the outputs of FCI and RFCI are
identical}\label{secrfci=fci}

We now specify graphical conditions on an underlying DAG $\mathcal
{G}=(\mathbf{V},\mathbf{E})$
with $\mathbf{V}=\mathbf{X}\ddcup \mathbf{L}\ddcup \mathbf{S}$ such
that the outputs of FCI and RFCI are identical (Theorem
\ref{thfci=rfci}). Moreover, if the outputs of RFCI and FCI are not
identical, we infer properties of edges that are present in the output of
RFCI but not in that of FCI (Theorem~\ref{thextraedgesrfci}).

The results in this section rely on the concept of inducing paths
\cite{SpirtesMeekRichardson99,VermaPearl90}, which we have
extended here:
%
\begin{defi}\label{defindpath}
Let $\mathcal{G}=(\mathbf{V},\mathbf{E})$ be a~DAG with
$\mathbf{V}=\mathbf{X}\ddcup \mathbf{L}\ddcup \mathbf{S}$ and
let~$\mathbf{Y}$
be a~subset of~$\mathbf{X}$ containing~$X_i$ and~$X_j$ with $X_i \neq
X_j$. A path $\pi$ between~$X_i$ and~$X_j$ is
called an \textit{inducing path relative to $\mathbf{Y}$ given
$\mathbf{S}$} if and only if
every member of $\mathbf{Y}\cup\mathbf{S}$ that is a~nonendpoint on
$\pi$ is a~collider on
$\pi$ and every collider on $\pi$ has a~descendant in $\{X_i,X_j\}\cup
\mathbf{S}$.
\end{defi}

We note that our Definition~\ref{defindpath} corresponds to the one in
\cite{SpirtesMeekRichardson99} if $\mathbf{Y}=\mathbf{X}$. The
existence of an inducing path in a~DAG is related to $d$-connection in
the following way. There is an inducing path between $X_i$ and $X_j$
relative to $\mathbf{Y}$ given
$\mathbf{S}$ if and only if $X_i$ and $X_j$ are not $d$-separated by
$(\mathbf{Y}^{\prime} \cup\mathbf{S})\setminus\{X_i,X_j \}$ for all
$\mathbf{Y}^{\prime}\subseteq\mathbf{Y}$ (see~\cite{SpirtesMeekRichardson99},
Lemma 9, page 243). The definition of an inducing path is monotone in
the following sense: if
$\mathbf{Y}_1 \subseteq\mathbf{Y}_2 \subseteq\mathbf{X}$ and there is
an inducing path between $X_i$
and $X_j$ relative to $\mathbf{Y}_2$ given $\mathbf{S}$, then there
also is an inducing path between $X_i$
and $X_j$ relative to $\mathbf{Y}_1$ given $\mathbf{S}$.

Consider a~pair of vertices $X_i,X_j \in\mathbf{X}$ in an underlying DAG
$\mathcal{G}$. We introduce the following shorthand notation. Let
Adj$(i,j)=\operatorname{adj}(\mathcal{C}_1,X_i)\setminus\{X_j\}$,
where $\mathcal{C}_1$ is the initial
skeleton after Step 1 of the algorithms.
Moreover, let Pds$(i,j)=\operatorname{pds}(\mathcal
{C}_2,X_i,X_j)\setminus\{X_j,X_i\}$,
where $\mathcal{C}_2$ is the graph resulting from Step
2 of the FCI algorithm. By definition, Pds$(k,i)\supseteq\operatorname
{Adj}(k,i)$
for any pair of vertices $X_k, X_i \in\mathbf{X}$. We now consider the
following three scenarios:
\begin{longlist}[(S3)]
\item[(S1)] There is an inducing path between $X_i$ and $X_j$ in
$\mathcal{G}$ relative to
Pds$(i,j)$ given $\mathbf{S}$, and there is an inducing path between
$X_i$ and $X_j$ relative to Pds$(j,i)$ given $\mathbf{S}$.
\item[(S2)] There is an inducing path between $X_i$ and $X_j$ in
$\mathcal{G}$ relative
to Adj$(i,j)$ given $\mathbf{S}$, and there is an inducing path between
$X_i$ and $X_j$ relative to Adj$(j,i)$ given\vadjust{\goodbreak} $\mathbf{S}$. Moreover,
there is no inducing path between $X_i$ and~$X_j$ in $\mathcal{G}$
relative to Pds$(i,j)$ given $\mathbf{S}$, or there is no inducing path
between~$X_i$ and~$X_j$ in~$\mathcal{G}$ relative to Pds$(j,i)$ given
$\mathbf{S}$.
\item[(S3)] There is no inducing path between $X_i$ and $X_j$ in
$\mathcal{G}$
relative to Adj$(i,j)$ given $\mathbf{S}$, or there is no inducing
path between $X_i$ and $X_j$ in $\mathcal{G}$ relative to Adj$(j,i)$
given $\mathbf{S}$.\vspace*{-2pt}
\end{longlist}

We now obtain the following theorem:\vspace*{-3pt}
%
\begin{theo}\label{thfci=rfci}
Assume that the distribution of $\mathbf{V}=\mathbf{X}\ddcup
\mathbf{L}\ddcup \mathbf{S}$ is faithful to an underlying DAG
$\mathcal{G}$. Then the output $\mathcal{C}^{\prime}$ of RFCI equals the
output $\mathcal{C}^{\prime\prime}$ of FCI if for every pair of vertices
$X_i$, $X_j$ in $\mathbf{X}$ either \textup{(S1)} holds or \textup{(S3)} holds. If
$\mathcal{C}^{\prime}\neq
\mathcal{C}^{\prime\prime}$, then the skeleton of $\mathcal{C}^{\prime}$
is a~strict superset of the skeleton of~$\mathcal{C}^{\prime\prime}$, and
Scenario \textup{(S2)} must hold for every pair of vertices that are adjacent
in~$\mathcal{C}'$ but not in $\mathcal{C}''$.\vspace*{-3pt}
\end{theo}

Scenario (S2) occurs if and only if (i) there is a~path $\pi(i,j)$
between~$X_i$ and~$X_j$ in the underlying DAG $\mathcal{G}$ that satisfies:
(c1) all
colliders on $\pi(i,j)$ have descendants in $\{X_i,X_j\}\cup\mathbf{S}$,
(c2) every member of $\operatorname{Adj}(i,j)\cup\mathbf{S}$ on $\pi
(i,j)$ is a~collider on
$\pi(i,j)$, (c3) there is a~member of $(\operatorname{Pds}(i,j) \cup
\operatorname{Pds}(j,i)) \setminus\operatorname{Adj}(i,j)$ on $\pi
(i,j)$ that is not a~collider on the path, and (ii) there is a~path $\pi(j,i)$ between $X_j$
and $X_i$
in the underlying DAG that satisfies conditions (c1)--(c3) above with
the roles of $i$ and $j$ reversed. In condition (c3), an equivalent
formulation is given by replacing $\operatorname{Pds}(i, j) \cup
\operatorname{Pds}(j, i)$
with $\mathbf{X} \setminus\{X_i,X_j\}$.\looseness=-1

To illustrate Theorem~\ref{thfci=rfci}, consider again Example
\ref{exRFCInotFCI} and the graphs in Figure~\ref{graphs}. The output of
RFCI for
the underlying DAG shown in Figure~\ref{graphs}(a) contains an edge
between $X_1$ and $X_5$, while the output of FCI does not. According to Theorem
\ref{thfci=rfci}, Scenario (S2) must hold for the vertices $X_1$ and~$X_5$.
Hence, there must exist paths $\pi(1,5)$ and $\pi(5,1)$ between
$X_1$ and $X_5$ in the underlying DAG that satisfy conditions (c1)--(c3)
above. This is indeed the case for the path $\pi= \pi(1,5) = \pi(5,1) =
\langle X_1,L_1,X_2,X_3,X_4,L_2,X_5\rangle$: (c1) there are
two colliders on $\pi$, $X_2$ and $X_4$, both with descendants in
$\{X_1,X_5\}$, (c2) all members of $\operatorname{Adj}(1,5) =
\operatorname{Adj}(5,1)=\{X_2,X_4\}$ on $\pi$ are colliders on the
path, and (c3) $X_3$ is
a~member of $(\operatorname{Pds}(1,5) \cup\operatorname{Pds}(5,1))
\setminus
\operatorname{Adj}(1,5) = (\operatorname{Pds}(5,1) \cup\operatorname
{Pds}(1,5))\setminus
\operatorname{Adj}(5,1)$ on $\pi$ and is a~noncollider on $\pi$.

To see that the occurrence of Scenario (S2) does not always lead to a~difference in the outputs of FCI and RFCI, we revisit Example
\ref{exRFCI=FCI} and the graphs in Figure
\ref{imprgraphs}. The same path $\pi$ as above satisfies conditions
(c1)--(c3) in the underlying DAG, but the outputs of FCI and RFCI are
identical (due to the extra tests in Step 2
of Algorithm~\ref{pseudorfci}). This illustrates that fulfillment of (S1)
or (S3) for every pair of vertices is not a~necessary condition for equality
of FCI and RFCI.

Finally, the following theorem establishes features of edges that are
present in an
RFCI-PAG but not in an FCI-PAG.\vspace*{-2pt}
%
\begin{theo}\label{thextraedgesrfci}
Assume that the distribution of $\mathbf{V}=\mathbf{X}\ddcup
\mathbf{L}\ddcup \mathbf{S}$ is faithful to an underlying DAG\vadjust{\goodbreak}
$\mathcal{G}$. If
there is an edge $X_i \twostar X_j$ in an RFCI-PAG for $\mathcal{G}$
that is not present in
an FCI-PAG for $\mathcal{G}$, then the following hold: \textup{(i)} $X_i \notin
\operatorname{an}(\mathcal{G},X_j\cup
\mathbf{S})$ and $X_j \notin\operatorname{an}(\mathcal{G},X_i \cup
\mathbf{S})$, and \textup{(ii)}
each edge mark of $X_i \twostar X_j$ in the RFCI-PAG is a~circle or an
arrowhead.
\end{theo}


\section{Consistency of FCI and RFCI in sparse high-dimensional
settings}\label{sechighdimconsistency}

Let $\mathcal{G}=(\mathbf{V},\mathbf{E})$ be a~DAG with $\mathbf
{V}=\mathbf{X} \ddcup \mathbf{L}
\ddcup \mathbf{S}$ and let $\mathcal{M}$ be the corresponding
unique MAG over
$\mathbf{X}$. We assume that we observe $n$ i.i.d. copies of $\mathbf
{W} = (W_1,\ldots,W_p) \sim(X_1|\mathbf{S}, \ldots, X_p|\mathbf{S})$.
To represent high-dimensional behavior, we let the DAG $\mathcal{G}$
and the number
of observed variables $p$ in $\mathbf{X}$ grow as a~function of the sample size, so that $p=p_n$, $\mathcal{G}=\mathcal
{G}_n$ and $\mathcal{M}=\mathcal{M}_n$. We do
not impose any restrictions on the number of latent and selection variables.
Throughout, we assume that $\mathbf{W}$ is multivariate Gaussian, so that
conditional independence is equivalent to zero partial correlation.

In Section~\ref{secsampleversionalgorithms}, we define the sample
versions of RFCI and the different versions of FCI. Sections
\ref{secconsistrfci} and~\ref{secconsistFCI} contain consistency
results for RFCI and FCI in sparse high-dimensional settings. The
conditions required for
consistency of RFCI are considerably weaker than those for FCI.

\subsection{Sample versions of RFCI and the different versions of
FCI}\label{secsampleversionalgorithms}

Let $\rho_{n;i,j\vert\mathbf{Y}}$ be the partial correlation between $W_i$
and $W_j$ in $\mathbf{W}$ given a~set $\mathbf{Y}\subseteq\mathbf{W}
\setminus
\{W_i,W_j\}$, and let $\hat{\rho}_{n;i,j\vert\mathbf{Y}}$ be the corresponding
sample partial correlation. We test if a~partial
correlation is equal to zero after applying Fisher's z-transform
defined as
$g(x)=\frac{1}{2}\log(\frac{1+x}{1-x})$. Thus, we consider
\[
\hat{z}_{n;i,j\vert\mathbf{Y}}=g(\hat{\rho}_{n;i,j\vert\mathbf{Y}})
\quad\mbox{and}\quad z_{n;i,j\vert\mathbf{Y}}=g(\rho_{n;i,j\vert\mathbf{Y}})
\]
and we reject the null-hypothesis $H_0(i,j\vert\mathbf{Y})\dvtx \rho
_{i,j\vert
\mathbf{Y}}=0$ against the two-sided alternative $H_A(i,j\vert
\mathbf{Y})\dvtx \rho_{i,j\vert\mathbf{Y}}\neq0$ at significance
level $\alpha$ if
%
\begin{equation} \label{decishypo}
\vert\hat{z}_{n;i,j \vert\mathbf{Y}}\vert> \Phi^{-1}(1-\alpha/2)
(n-\vert\mathbf{Y} \vert-3)^{-1/2},
\end{equation}
where $\Phi(\cdot)$ denotes the cumulative distribution function of a~standard normal random variable. (We assume $n>|\mathbf{Y}|+3$.)

Sample versions of RFCI and the different versions of FCI can be obtained
by simply adapting all steps with conditional independence decisions
as follows: $X_i$~and $X_j$ are judged to be conditionally independent given
$\mathbf{Y}^{\prime}\cup\mathbf{S}$ for $\mathbf{Y}^{\prime}\subseteq
\mathbf{X}\setminus\{X_i,X_j\}$ if and
only if $\vert\hat{z}_{n;i,j \vert\mathbf{Y}}\vert\leq
\Phi^{-1}(1-\alpha/2) (n-\vert\mathbf{Y} \vert-3)^{-1/2}$ for $\mathbf{Y}
\sim\mathbf{Y}' | \mathbf{S}$. The parameter $\alpha$ is used for many
tests, and plays the role of a~tuning parameter.

\subsection{Consistency of RFCI}\label{secconsistrfci}

We impose the following assumptions:
\begin{longlist}[(A5)]
\item[(A1)] The distribution of $\mathbf{W}$ is faithful to the
underlying causal MAG $\mathcal{M}_n$ for all
$n$.
\item[(A2)] The number of variables in $\mathbf{X}$, denoted by
$p_n$, satisfies
$p_n=O(n^a)$ for some $0 \leq a~< \infty$.\vadjust{\goodbreak}
\item[(A3)] The maximum size of the adjacency sets after Step 1 of the
oracle RFCI algorithm, denoted by $q_n=\max_{1\leq i \leq p_n}(\vert
{\operatorname{adj}}(\mathcal{C}_1,X_i)\vert)$, where $\mathcal{C}_1$ is
the skeleton\vspace*{1pt} after Step
1, satisfies $q_n=O(n^{1-b})$ for some $0 < b \leq1$.
\item[(A4)] The distribution of $\mathbf{W}$ is multivariate Gaussian.
\item[(A5)] The partial correlations satisfy the following lower and
upper bound for
all $W_i,W_j \in\{W_1,\ldots, W_{p_n}\}$ and $\mathbf{Y} \subseteq
\{W_1,\ldots,W_{p_n}\}\setminus\{W_i,W_j\}$ with
$\vert\mathbf{Y}\vert\leq q_n$:
\begin{eqnarray*}
\inf\{\vert\rho_{n;i,j\vert\mathbf{Y}}\vert\dvtx \rho_{n;i,j\vert
\mathbf{Y}}\neq0 \} &\geq& c_n, \\
\sup\{\vert\rho_{n;i,j\vert\mathbf{Y}}\vert\dvtx i \neq j\} &\leq& M<1,
\end{eqnarray*}
where $c_n^{-1}=O(n^d)$ for some $0\leq d<b/2$ with $b$ from (A3).
\end{longlist}
Assumption (A2) allows the number of variables to grow as any
polynomial of the sample size, representing a~high-dimensional
setting. Assumption (A3) is a~sparseness assumption, and poses a~bound
on the growth of the maximum size of the
adjacency sets in the graph resulting from Step 1 of the oracle RFCI algorithm.
The upper bound in assumption (A5) excludes sequences of models in
which the partial
correlations tend to 1, hence avoiding identifiability problems. The
lower bound in assumption (A5) requires the nonzero
partial correlations to be outside of the $n^{-b/2}$ range, with $b$
as in assumption (A3). This condition is similar to assumption 5 in
\cite{MeinshausenBuehlmann06} and condition (8) in~\cite{ZhaoYu06}.

The similarities between our assumptions and the assumptions of
\cite{KalischBuehlmann07a} for consistency of the PC algorithm are
evident. The main differences are that our assumption (A3) concerns
the skeleton after Step 1 of the oracle RFCI algorithm instead of the
underlying DAG, and that our assumptions (A1) and (A4)--(A5) concern
the distribution of $\mathbf{W}$ instead of $\mathbf{X}$.
%
\begin{theo}\label{consrfci}
Assume \textup{(A1)--(A5)}. Denote by $\hat{\mathcal{C}}_n(\alpha_n)$ the output
of the sample version of the RFCI algorithm and by
$\mathcal{C}_n^{\prime}$ the oracle version of the RFCI algorithm.
Then there exists a~sequence $\alpha_n
\rightarrow0$ $(n \rightarrow\infty)$ and a~constant $0<C<\infty$
such that
\[
\Prob[\hat{\mathcal{C}}_{n}(\alpha_n)=\mathcal{C}_n^{\prime}] \ge
1-O(\exp(-Cn^{1-2d})) \rightarrow1
\qquad\mbox{as } n \rightarrow\infty,
\]
where $d>0$ is as in (A5).
\end{theo}

One such sequence for $\alpha_n$ is
$\alpha_n=2(1-\Phi(n^{1/2}c_n/2))$, where $c_n$ is the lower bound
in (A5) (which depends on the unknown data distribution).


\subsection{Consistency of FCI}\label{secconsistFCI}

Assume (A1)--(A5) of Section~\ref{secconsistrfci},
but replace (A3) by (A3$'$):
\begin{longlist}[(A3$'$)]
\item[(A3$'$)] The maximum size of the Possible-D-SEP sets in Step 3
of the oracle FCI algorithm, denoted by $r_n=\max_{1\leq i \leq
p_n}(\vert{\operatorname{pds}}(\mathcal{C}_2,X_i,\cdot)\vert)$, where
$\mathcal{C}_2$ is\vspace*{1pt} the graph
resulting from Step 2, satisfies $r_n=O(n^{1-b})$ for some $0 < b
\leq1$.\vadjust{\goodbreak}
\end{longlist}

Assumption (A3$'$) is stronger than assumption (A3), since the skeleton after
Step 1 of the RFCI algorithm is identical to the skeleton after
Step 2 of the FCI algorithm, and since the adjacency set is contained in
Possible-D-SEP by definition. (In fact, one can construct sequences
of graphs in which the maximum size of the adjacency sets is fixed,
but the maximum size of the Possible-D-SEP sets grows linearly with
the number of vertices.) The stricter assumption (A3$'$) is needed
for the additional conditional independence tests in Step 3 of the FCI
algorithm.
%
\begin{theo}\label{constheo}
Assume \textup{(A1)--(A5)} with \textup{(A3$'$)} instead of \textup{(A3)}.
Consider one of the sample
versions of FCI, FCI$_{\mathrm{path}}$, CFCI, CFCI$_{\mathrm{path}}$,
SCFCI or
SCFCI$_{\mathrm{path}}$, and denote its output by
$\mathcal{C}^{*}_{n}(\alpha_n)$. Denote the true underlying FCI-PAG by
$\mathcal{C}_n$. Then there
exists a~sequence $\alpha_n \rightarrow0$ $(n \rightarrow\infty)$ and
a~constant $0<C<\infty$ such that
\[
\Prob[\mathcal{C}^{*}_{n}(\alpha_n)=\mathcal{C}_n] \ge1-O(\exp
(-Cn^{1-2d})) \rightarrow1\qquad
\mbox{as } n \rightarrow\infty,
\]
where $d>0$ is as in \textup{(A5)}.
\end{theo}

As before, one such sequence for $\alpha_n$ is
$\alpha_n=2(1-\Phi(n^{1/2}c_n/2))$, where $c_n$ is the lower bound in (A5).


\section{Numerical examples}\label{secsimulations}

In this section we compare the performance of RFCI and different
versions of FCI and
Anytime FCI in simulation studies, considering both the computing time
and the
estimation performance. Since Anytime FCI requires an additional tuning
parameter (see~\cite{Spirtes01-anytime} and Section 3 of
\cite{ColomboEtAl12-supp}), we cannot compare it directly. We therefore
define a~slight modification, called Adaptive Anytime FCI (AAFCI), where
this tuning parameter is set adaptively (see Section 3
of~\cite{ColomboEtAl12-supp}).
Our proposed modifications of FCI (see Section~\ref{secfci})
can also be applied to AAFCI, leading to the following algorithms:
AAFCI$_{\mathrm{path}}$, CAAFCI, CAAFCI$_{\mathrm{path}}$, SCAAFCI
and SCAAFCI$_{\mathrm{path}}$.

The remainder of this section is organized as follows.
The simulation setup is described in Section
\ref{secsimulationsetup}. Section~\ref{secestimationperformance}
shows
that the estimation performances of RFCI and all versions of FCI and AAFCI
are very similar. Section~\ref{seccomputingtime} shows that our
adaptations of FCI and AAFCI can reduce the computation time significantly
for graphs of moderate size, but that RFCI is the
only feasible algorithm for large graphs.

\subsection{Simulation setup}\label{secsimulationsetup}

We use the following procedure to generate a~random DAG with a~given number
of vertices $p'$ and expected neighborhood size $E(N)$. First, we generate
a~random adjacency matrix $A$ with independent realizations of
Bernoulli$(E(N)/(p'-1))$ random variables in the lower triangle of the
matrix and zeroes in the remaining entries. Next, we replace the ones in
$A$ by independent realizations of a~Uniform($[0.1,1]$) random
variable. A
nonzero entry $A_{ij}$ can be interpreted as an edge from $X_j$ to $X_i$
with ``strength'' $A_{ij}$, in the sense that $X_1,\ldots,X_{p'}$ can be
generated as follows: $X_1 = \varepsilon_1$ and $X_i =
\sum_{r=1}^{i-1}A_{ir}X_r + \varepsilon_i$ for $i=2,\ldots,p'$, where
$\varepsilon_1,\ldots,\varepsilon_{p'}$ are mutually independent
$\mathcal{N}(0,1)$ random variables. The variables $X_1,\ldots, X_{p'}$
then have a~multivariate Gaussian distribution with mean zero and
covariance matrix $\Sigma' = (\mathbh{1}-A)^{-1}(\mathbh{1}-A)^{-T}$, where
$\mathbh{1}$ is the $p' \times p'$ identity matrix.

To assess the impact of latent variables, we randomly define half of the
variables that have no parents and at least two children to be latent
(we do not consider selection variables). We restrict ourselves to
variables that have no parents and at least two
children, since these are particularly difficult for RFCI in the sense
that they are likely to
satisfy Scenario (S2) in Section~\ref{secrfci=fci}. Throughout, we let $p$
denote the number of observed variables.

We consider the oracle versions of RFCI and FCI$_{\mathrm{path}}$ (note that
the outputs of FCI$_{\mathrm{path}}$ and FCI are identical in the
oracle versions), and
the sample versions of RFCI, (AA)FCI, (AA)FCI$_{\mathrm{path}}$,
C(AA)FCI, C(AA)FCI$_{\mathrm{path}}$ and SC(AA)FCI$_{\mathrm{path}}$.
In all plots (AA)FCI$_{\mathrm{path}}$ is abbreviated as (AA)FCIp.
Let~$\Sigma$ be the $p \times p$ matrix that is obtained from $\Sigma'$ by
deleting the rows and columns that correspond to latent variables. The
oracle versions of the algorithms use $\Sigma$ as input, and the sample
versions of the algorithms use simulated data from a~$N_p(0,\Sigma)$
distribution as input.

The simulations were performed on an AMD Opteron (tm) Quad Core Processor
8380 with 2.5 GHz and 2 GB RAM on Linux using R 2.11.0.

\subsection{Estimation performance}\label{secestimationperformance}

We first investigated the difference between the oracle versions of
RFCI and
FCI$_{\mathrm{path}}$, using simulation settings $p' \in\{15,20, 25\}$
and $E(N)=2$. For each combination of these parameters, we generated
1000 DAGs, where the average number of observed variables was $p \approx
\{14, 18, 23\}$ (rounded to the nearest integer). For each simulated graph,
we assessed whether the outputs of FCI$_{\mathrm{path}}$ and RFCI were
different, and if this was the case, we counted the number of additional
edges in the output of RFCI when compared to that of FCI. For $p'=15$, $p'=20$
and $p'=25$, there were 0, 1 and 5 of the 1000 DAGs that gave different
results, and whenever there was a~difference, the output of RFCI had a~single additional edge.
Hence, for these simulation settings, the oracle versions of
FCI$_{\mathrm{path}}$ and RFCI were almost always identical, and if
there was a~difference, the difference was very small.

Next, we investigated the performance of the sample versions of RFCI
and our adaptations
of FCI and AAFCI, considering the number of differences
in the output when compared to the true FCI-PAG. We used two simulation
settings: small-scale and large-scale.

The small-scale simulation setting is as follows. For
each value of $p' \in\{10,15$, $20,25,30\}$, we generated 50 random DAGs
with $E(N)=2$,
where the average number of observed variables was $p \approx\{9, 14, 18,
23, 27\}$. For each such DAG, we generated a~data set of size $n=1000$ and
ran RFCI, (AA)FCI, (AA)FCI$_{\mathrm{path}}$, C(AA)FCI and
SC(AA)FCI$_{\mathrm{path}}$
with tuning parameter $\alpha=0.01$.

\begin{figure}
\begin{tabular}{@{}cc@{}}

\includegraphics{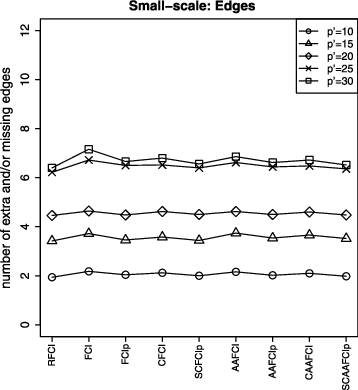}
 & \includegraphics{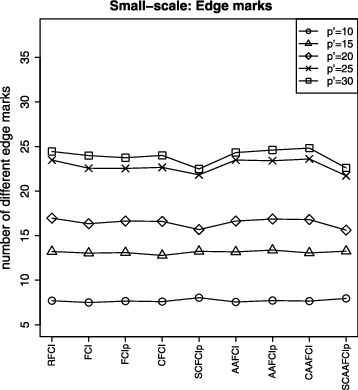} \\
(a) & (b)
\end{tabular}
\caption{Estimation performance of the sample versions of RFCI and the
different versions of FCI and AAFCI in the small-scale setting, when
compared to the true underlying FCI-PAG. The simulation settings
were $E(N)=2$, $n=1000$ and $\alpha=0.01$.
\textup{(a)} Average number of missing or extra edges over 50 replicates;
\textup{(b)} average number of different edge marks over 50 replicates.}
\label{figperf}
\end{figure}

Figure~\ref{figperf} shows the results for the
small-scale setting. Figure~\ref{figperf}(a) shows the average number
of missing or extra edges
over the 50 replicates, and we see that this number was virtually identical
for all algorithms. Figure~\ref{figperf}(b) shows the average number of
different edge marks over the 50 replicates. We again see that all
algorithms performed similarly. We note that the conservative and
superconservative
adaptations of the algorithms yield slightly better edge orientations than
the standard versions for larger graphs.

%

%
\begin{figure}
\begin{tabular}{@{}cc@{}}

\includegraphics{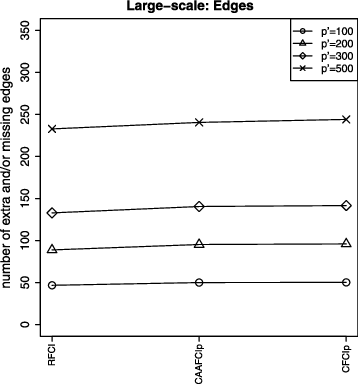}
 & \includegraphics{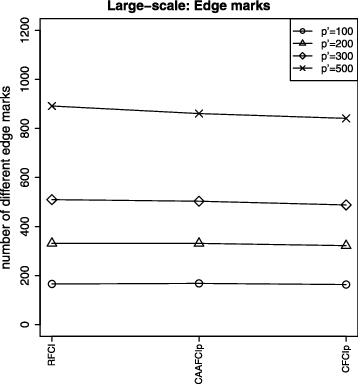} \\
(a) & (b)\vspace*{-3pt}
\end{tabular}
\caption{Estimation performance of the sample versions of RFCI and the
fastest versions
of FCI and AAFCI in the large-scale setting, when compared to the true
underlying
FCI-PAG. The simulation settings were $E(N)=3$, $n=1000$ and $\alpha=0.01$.
\textup{(a)} Average number of missing or extra edges over $91$
replicates (see text);
\textup{(b)} average number of different edge marks over $91$
replicates (see text).}
\label{figperflarge}\vspace*{-3pt}
\end{figure}

The large-scale simulation setting is as follows. For
each value of $p'\in\{100$, $200,300,500\}$ we generated 100 random DAGs
with $E(N)=3$, where the average number of observed
variables was $p \approx\{90, 180, 271, 452\}$. For each DAG, we
generated a~data
set of size $n=1000$, and ran RFCI, CFCI$_{\mathrm{path}}$ and
CAAFCI$_{\mathrm{path}}$ [the other versions of (AA)FCI were
computationally infeasible]
using tuning parameter $\alpha=0.01$. To ensure reasonable computing times,
we terminated an algorithm for a~graph if it was not finished after
eight hours.
For CFCI$_{\mathrm{path}}$, termination occurred five times for
$p'=300$ and nine
times for $p'=500$. One of the latter nine graphs also led to
termination of CAAFCI$_{\mathrm{path}}$.
To ensure comparability we deleted any run which did not complete for
all
algorithms and computed the average number of missing or extra edges
[see Figure~\ref{figperflarge}(a)] and the average number of different
edge marks
[see Figure~\ref{figperflarge}(b)] over the 91 remaining runs.
We again see that all algorithms performed similarly.

\subsection{Computing time} \label{seccomputingtime}

\begin{figure}
\begin{tabular}{@{}cc@{}}

\includegraphics{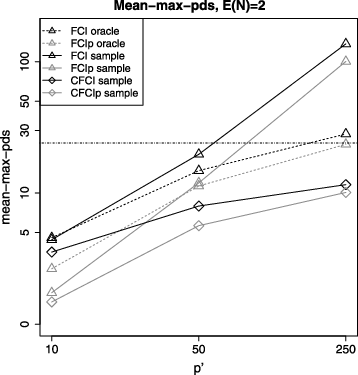}
 & \includegraphics{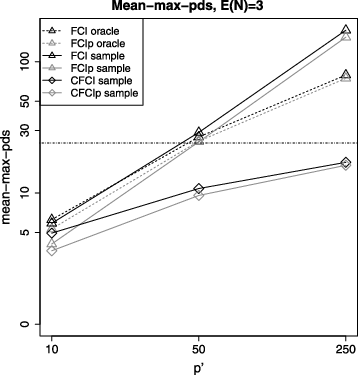}
\end{tabular}
\caption{A plot of mean-max-pds (see text) versus $p'$, where both axes
are drawn in
log scale. The horizontal line at mean-max-pds${} = {}$24 indicates an upper bound
that still yields a~feasible running time of the algorithms.}
\label{figsizepds}\vspace*{-3pt}
\end{figure}

We first compared the size of the Possible-D-SEP sets in the different versions
of FCI, since this is the most important factor for the computing time of
these algorithms. In particular, if the size of Possible-D-SEP is, say, 25
vertices or more, it becomes computationally infeasible to consider all
of its subsets.
For all combinations of $p' \in\{10,50,250\}$ and $E(N) \in\{2,3\}$, we
generated 100 random graphs and ran the oracle version of FCI and
FCI$_{\mathrm{path}}$ and the sample versions of FCI, FCI$_{\mathrm
{path}}$, CFCI
and CFCI$_{\mathrm{path}}$. The average number of observed
variables was $p \approx\{9, 46, 230\}$ for $E(N)=2$ and $p \approx\{
9, 45, 226\}$ for
$E(N)=3$. For the sample versions of the algorithms we used sample size
$n=1000$ and tuning parameter $\alpha=0.01$.
For each simulated graph and each algorithm we computed the maximum
size of
the Possible-D-SEP sets over all vertices in the graph. We averaged these
numbers over the 100 replicates, and denoted the result by
mean-max-pds. The results are shown in Figure~\ref{figsizepds}. We see
that the new definition of $\mathrm{pds}_{\mathrm{path}}$ (see Definition
\ref{defpdspath} used in algorithm FCI$_{\mathrm{path}}$ and
CFCI$_{\mathrm{path}}$) reduced mean-max-pds
slightly, while the conservative adaptations of the sample versions of the\vadjust{\goodbreak}
algorithms reduced it drastically. These results are also relevant for
the different versions of AAFCI, since AAFCI
considers all subsets of Possible D-SEP up to a~certain size. This
again becomes infeasible
if Possible D-SEP is large.

%

Next, we investigated the computing time of the sample version of RFCI and
modifications of FCI and AAFCI under the same simulation settings as in
Section~\ref{secestimationperformance}.

%
\begin{figure}
\begin{tabular}{@{}cc@{}}

\includegraphics{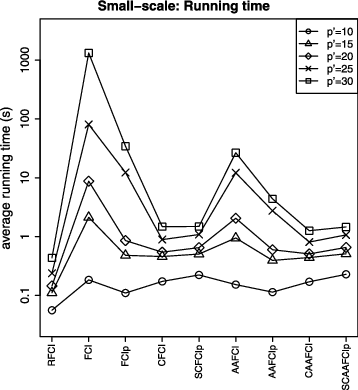}
 & \includegraphics{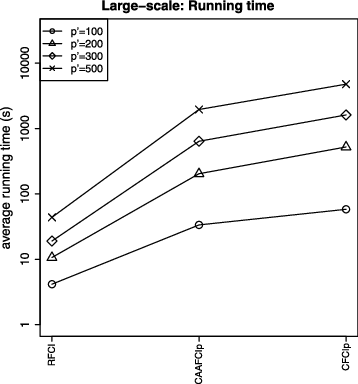} \\
(a) & (b)
\end{tabular}
\caption{Running time of the sample versions of the algorithms, using
simulation settings $n=1000$ and $\alpha=0.01$, where the y-axes are
drawn in log scale. \textup{(a)} Average running time in seconds of each
algorithm over $50$ replicates, using $E(N)=2$; \textup{(b)}
average running time in seconds of each algorithm over $91$
replicates (see text), using $E(N)=3$.}
\label{figperftime}
\end{figure}

Figure~\ref{figperftime}(a) shows the average running times over the 50
replicates in
the small-scale setting. We see that RFCI was fastest for all parameter
settings, while
the standard version of FCI was slowest for all settings with $p' \geq
15$. Our new adaptations of FCI and AAFCI reduced the running time of
FCI and AAFCI significantly, which is in correspondence with the
reduction in
mean-max-pds that we saw in Figure~\ref{figsizepds}.

Figure~\ref{figperftime}(b) shows the average running times
over the 91 fastest runs in the large-scale setting.
We see that RFCI is the only algorithm that is computationally feasible
for large graphs: for
$p'=500$ RFCI took about 40 seconds, while the fastest modifications of FCI
took about 10,000 seconds. These results can be explained by the fact
that Steps 2 and 3 in the RFCI algorithm
only involve local tests (conditioning on subsets of the adjacency set
of a~vertex), while
Step 3 of (AA)FCI considers subsets of the Possible D-SEP sets, which
can be large
even for sparse graphs (see Figure~\ref{figsizepds}).

\section{Discussion}

In this paper, we introduce a~new algorithm for learning PAGs, called the
Really Fast Causal Inference (RFCI) algorithm. RFCI uses fewer conditional
independence tests than the existing FCI algorithm, and its tests condition
on a~smaller number of variables.\vadjust{\goodbreak}

The output of RFCI can be interpreted as the output of FCI, with the only
difference that the presence of an edge has a~weaker meaning. In
particular, the interpretation
of tails and arrowheads is identical for both algorithms. In this sense
the RFCI algorithm is similar
to the Anytime FCI algorithm of~\cite{Spirtes01-anytime}.

We describe a~class of graphs where the outputs of FCI and RFCI are
identical, and show that differences between the two algorithms are
caused by very special structures in the underlying DAG.
We confirm this finding in simulation studies that show that
differences between the oracle versions of RFCI and FCI are very rare.

We prove consistency of FCI and RFCI in sparse high-dimensional
settings. The sparsity conditions needed for consistency of RFCI are
considerably weaker than those needed for FCI, due to the lower
computational complexity of the RFCI algorithm.

We compare RFCI with several modifications of (Anytime) FCI in simulation
studies. We show that all algorithms perform similarly in terms of
estimation, and that RFCI is the only algorithm that is computationally
feasible for high-dimensional sparse graphs.

We envision several possible uses of RFCI. First, it could be used in
addition to the PC algorithm to assess the potential impact
of the existence of latent or selection variables. Second, it could be
used as a~building block for an IDA-like method
\cite{MaathuisColomboKalischBuhlmann10,MaathuisKalischBuehlmann09}
to obtain bounds on causal effects based on observational data that is
faithful to an \textit{unknown} underlying causal graph \textit{with
arbitrarily many latent and selection variables}.
In order to achieve the latter, we plan to build on the work of
\cite{RichardsonSpirtes03,Zhang08-causal-reasoning-ancestral-graphs}, who
made a~start with the study of causal reasoning for ancestral graphs.
Other interesting open problems include investigating which RFCI-PAGs can
only correspond to a~single Markov equivalence class, 
and investigating completeness of the RFCI algorithm, that is,
investigating whether the edge marks in the output of RFCI are
maximally informative.

\begin{supplement}[id=suppA]
\stitle{Supplement to ``Learning high-dimensional directed acyclic
graphs with latent and selection variables''}
\slink[doi]{10.1214/11-AOS940SUPP} 
\sdatatype{.pdf}
\sfilename{aos940\_supp.pdf}
\sdescription{All proofs, a~description of the Adaptive Anytime FCI
algorithm, pseudocodes, and two additional examples can be found in the
supplementary document~\cite{ColomboEtAl12-supp}.}
\end{supplement}

%

\printaddresses


\begin{thebibliography}{26}

\bibitem{Aho74}
%
\begin{bbook}[author]
\bauthor{\bsnm{Aho},~\bfnm{Alfred}\binits{A.}},
\bauthor{\bsnm{Hopcroft},~\bfnm{John}\binits{J.}} \AND
\bauthor{\bsnm{Ullman},~\bfnm{Jeffrey~D.}\binits{J.~D.}}
(\byear{1974}).
\btitle{The Design and Analysis of Computer Algorithms}.
\bpublisher{Addison-Wesley}, \baddress{Boston, MA}.
\bptok{imsref}%
\end{bbook}
%
\endbibitem

\bibitem{Ali09equiv}
%
\begin{barticle}[mr]
\bauthor{\bsnm{Ali},~\bfnm{R.~Ayesha}\binits{R.~A.}},
\bauthor{\bsnm{Richardson},~\bfnm{Thomas~S.}\binits{T.~S.}} \AND
\bauthor{\bsnm{Spirtes},~\bfnm{Peter}\binits{P.}}
(\byear{2009}).
\btitle{Markov equivalence for ancestral graphs}.
\bjournal{Ann. Statist.}
\bvolume{37}
\bpages{2808--2837}.
\bid{doi={10.1214/08-AOS626}, issn={0090-5364}, mr={2541448}}
\bptok{imsref}%
\end{barticle}
%
\endbibitem

\bibitem{AndersonEtAll97}
%
\begin{barticle}[mr]
\bauthor{\bsnm{Andersson},~\bfnm{Steen~A.}\binits{S.~A.}},
\bauthor{\bsnm{Madigan},~\bfnm{David}\binits{D.}} \AND
\bauthor{\bsnm{Perlman},~\bfnm{Michael~D.}\binits{M.~D.}}
(\byear{1997}).
\btitle{A characterization of {M}arkov equivalence classes for acyclic
digraphs}.
\bjournal{Ann. Statist.}
\bvolume{25}
\bpages{505--541}.
\bid{doi={10.1214/aos/1031833662}, issn={0090-5364}, mr={1439312}}
\bptok{imsref}%
\end{barticle}
%
\endbibitem

\bibitem{Chickering02}
%
\begin{barticle}[mr]
\bauthor{\bsnm{Chickering},~\bfnm{David~Maxwell}\binits{D.~M.}}
(\byear{2002}).
\btitle{Learning equivalence classes of {B}ayesian-network structures}.
\bjournal{J.~Mach. Learn. Res.}
\bvolume{2}
\bpages{445--498}.
\bid{doi={10.1162/153244302760200696}, issn={1532-4435}, mr={1929415}}
\bptok{imsref}%
\end{barticle}
%
\endbibitem

\bibitem{ColomboEtAl12-supp}
%
\begin{bmisc}[author]
\bauthor{\bsnm{Colombo},~\bfnm{Diego}\binits{D.}},
\bauthor{\bsnm{Maathuis},~\bfnm{Marloes~H.}\binits{M.~H.}},
\bauthor{\bsnm{Kalisch},~\bfnm{Markus}\binits{M.}} \AND
\bauthor{\bsnm{Richardson},~\bfnm{Thomas~S.}\binits{T.~S.}}
(\byear{2012}).
\bhowpublished{Supplement to ``Learning high-dimensional directed acyclic
graphs with
latent and selection variables.''
DOI:\doiurl{10.1214/11-AOS940SUPP}.}
\bptok{imsref}%
\end{bmisc}
%
\endbibitem

\bibitem{Cooper95}
%
\begin{binproceedings}[author]
\bauthor{\bsnm{Cooper},~\bfnm{G.}\binits{G.}}
(\byear{1995}).
\btitle{Causal discovery from data in the presence of selection bias}.
In \bbooktitle{Preliminary Papers of the Fifth International Workshop on
Artificial Intelligence and Statistics}
(\beditor{D. Fisher}, ed.)
\bpages{140--150}.
\bptok{imsref}%
\end{binproceedings}
%
\endbibitem

\bibitem{Dawid80}
%
\begin{barticle}[mr]
\bauthor{\bsnm{Dawid},~\bfnm{A.~Philip}\binits{A.~P.}}
(\byear{1980}).
\btitle{Conditional independence for statistical operations}.
\bjournal{Ann. Statist.}
\bvolume{8}
\bpages{598--617}.
\bid{issn={0090-5364}, mr={0568723}}
\bptok{imsref}%
\end{barticle}
%
\endbibitem

\bibitem{KalischBuehlmann07a}
%
\begin{barticle}[author]
\bauthor{\bsnm{Kalisch},~\bfnm{M.}\binits{M.}} \AND
\bauthor{\bsnm{B{\"u}hlmann},~\bfnm{P.}\binits{P.}}
(\byear{2007}).
\btitle{Estimating high-dimensional directed acyclic graphs with the
{PC}-algorithm}.
\bjournal{J. Mach. Learn. Res.}
\bvolume{8}
\bpages{613--636}.
\bptok{imsref}%
\end{barticle}
%
\endbibitem

\bibitem{KalischEtAl11}
%
\begin{bmisc}[author]
\bauthor{\bsnm{Kalisch},~\bfnm{M.}\binits{M.}},
\bauthor{\bsnm{M{\"a}chler},~\bfnm{M.}\binits{M.}},
\bauthor{\bsnm{Colombo},~\bfnm{D.}\binits{D.}},
\bauthor{\bsnm{Maathuis},~\bfnm{M.~H.}\binits{M.~H.}} \AND
\bauthor{\bsnm{B{\"u}hlmann},~\bfnm{P.}\binits{P.}}
(\byear{2012}).
\bhowpublished{Causal inference using graphical models with the R package pcalg.
\textit{J.~Statist. Software}. To appear.}
\bptok{imsref}%
\end{bmisc}
%
\endbibitem

\bibitem{MaathuisColomboKalischBuhlmann10}
%
\begin{barticle}[pbm]
\bauthor{\bsnm{Maathuis},~\bfnm{Marloes~H.}\binits{M.~H.}},
\bauthor{\bsnm{Colombo},~\bfnm{Diego}\binits{D.}},
\bauthor{\bsnm{Kalisch},~\bfnm{Markus}\binits{M.}} \AND
\bauthor{\bsnm{B{\"{u}}hlmann},~\bfnm{Peter}\binits{P.}}
(\byear{2010}).
\btitle{Predicting causal effects in large-scale systems from observational
data}.
\bjournal{Nat. Methods}
\bvolume{7}
\bpages{247--248}.
\bid{doi={10.1038/nmeth0410-247}, issn={1548-7105}, pii={nmeth0410-247},
pmid={20354511}}
\bptok{imsref}%
\end{barticle}
%
\endbibitem

\bibitem{MaathuisKalischBuehlmann09}
%
\begin{barticle}[mr]
\bauthor{\bsnm{Maathuis},~\bfnm{Marloes~H.}\binits{M.~H.}},
\bauthor{\bsnm{Kalisch},~\bfnm{Markus}\binits{M.}} \AND
\bauthor{\bsnm{B{\"u}hlmann},~\bfnm{Peter}\binits{P.}}
(\byear{2009}).
\btitle{Estimating high-dimensional intervention effects from observational
data}.
\bjournal{Ann. Statist.}
\bvolume{37}
\bpages{3133--3164}.
\bid{doi={10.1214/09-AOS685}, issn={0090-5364}, mr={2549555}}
\bptok{imsref}%
\end{barticle}
%
\endbibitem

\bibitem{MeinshausenBuehlmann06}
%
\begin{barticle}[mr]
\bauthor{\bsnm{Meinshausen},~\bfnm{Nicolai}\binits{N.}} \AND
\bauthor{\bsnm{B{\"u}hlmann},~\bfnm{Peter}\binits{P.}}
(\byear{2006}).
\btitle{High-dimensional graphs and variable selection with the lasso}.
\bjournal{Ann. Statist.}
\bvolume{34}
\bpages{1436--1462}.
\bid{doi={10.1214/009053606000000281}, issn={0090-5364}, mr={2278363}}
\bptok{imsref}%
\end{barticle}
%
\endbibitem

\bibitem{Pearl00}
%
\begin{bbook}[mr]
\bauthor{\bsnm{Pearl},~\bfnm{Judea}\binits{J.}}
(\byear{2000}).
\btitle{Causality. Models, Reasoning, and Inference}.
\bpublisher{Cambridge Univ. Press}, \baddress{Cambridge}.
\bid{mr={1744773}}
\bptok{imsref}%
\end{bbook}
%
\endbibitem

\bibitem{Pearl09}
%
\begin{barticle}[mr]
\bauthor{\bsnm{Pearl},~\bfnm{Judea}\binits{J.}}
(\byear{2009}).
\btitle{Causal inference in statistics: An overview}.
\bjournal{Stat. Surv.}
\bvolume{3}
\bpages{96--146}.
\bid{doi={10.1214/09-SS057}, issn={1935-7516}, mr={2545291}}
\bptok{imsref}%
\end{barticle}
%
\endbibitem

\bibitem{RamseyZhangSpirtes06}
%
\begin{binproceedings}[author]
\bauthor{\bsnm{Ramsey},~\bfnm{Joseph}\binits{J.}},
\bauthor{\bsnm{Zhang},~\bfnm{Jiji}\binits{J.}} \AND
\bauthor{\bsnm{Spirtes},~\bfnm{Peter}\binits{P.}}
(\byear{2006}).
\btitle{Adjacency-faithfulness and conservative causal inference}.
In \bbooktitle{Proceedings of the 22nd Annual Conference on Uncertainty in
Artificial Intelligence}.
\bpublisher{AUAI Press}, \baddress{Arlington, VA}.
\bptok{imsref}%
\end{binproceedings}
%
\endbibitem

\bibitem{RichardsonSpirtes02}
%
\begin{barticle}[mr]
\bauthor{\bsnm{Richardson},~\bfnm{Thomas}\binits{T.}} \AND
\bauthor{\bsnm{Spirtes},~\bfnm{Peter}\binits{P.}}
(\byear{2002}).
\btitle{Ancestral graph {M}arkov models}.
\bjournal{Ann. Statist.}
\bvolume{30}
\bpages{962--1030}.
\bid{doi={10.1214/aos/1031689015}, issn={0090-5364}, mr={1926166}}
\bptok{imsref}%
\end{barticle}
%
\endbibitem

\bibitem{RichardsonSpirtes03}
%
\begin{bincollection}[mr]
\bauthor{\bsnm{Richardson},~\bfnm{Thomas~S.}\binits{T.~S.}} \AND
\bauthor{\bsnm{Spirtes},~\bfnm{Peter}\binits{P.}}
(\byear{2003}).
\btitle{Causal inference via ancestral graph models}.
In \bbooktitle{Highly Structured Stochastic Systems}.
\bseries{Oxford Statistical Science Series}
\bvolume{27}
\bpages{83--113}.
\bpublisher{Oxford Univ. Press}, \baddress{Oxford}.
\bid{mr={2082407}}
\bptok{imsref}%
\end{bincollection}
%
\endbibitem

\bibitem{RobinsEtAl00}
%
\begin{barticle}[pbm]
\bauthor{\bsnm{Robins},~\bfnm{J.~M.}\binits{J.~M.}},
\bauthor{\bsnm{Hern{\'{a}}n},~\bfnm{M.~A.}\binits{M.~A.}} \AND
\bauthor{\bsnm{Brumback},~\bfnm{B.}\binits{B.}}
(\byear{2000}).
\btitle{Marginal structural models and causal inference in epidemiology}.
\bjournal{Epidemiology}
\bvolume{11}
\bpages{550--560}.
\bid{issn={1044-3983}, pmid={10955408}}
\bptok{imsref}%
\end{barticle}
%
\endbibitem

\bibitem{Spirtes01-anytime}
%
\begin{binproceedings}[author]
\bauthor{\bsnm{Spirtes},~\bfnm{Peter}\binits{P.}}
(\byear{2001}).
\btitle{An anytime algorithm for causal inference}.
In \bbooktitle{Proc. of the Eighth International Workshop on Artificial
Intelligence and Statistics}
\bpages{213--221}.
\bpublisher{Morgan Kaufmann}, \baddress{San Francisco}.
\bptok{imsref}%
\end{binproceedings}
%
\endbibitem

\bibitem{SpirtesEtAl00}
%
\begin{bbook}[mr]
\bauthor{\bsnm{Spirtes},~\bfnm{Peter}\binits{P.}},
\bauthor{\bsnm{Glymour},~\bfnm{Clark}\binits{C.}} \AND
\bauthor{\bsnm{Scheines},~\bfnm{Richard}\binits{R.}}
(\byear{2000}).
\btitle{Causation, Prediction, and Search},
\bedition{2nd} ed.
\bpublisher{MIT Press}, \baddress{Cambridge, MA}.
\bid{mr={1815675}}
\bptok{imsref}%
\end{bbook}
%
\endbibitem

\bibitem{SpirtesMeekRichardson99}
%
\begin{bincollection}[mr]
\bauthor{\bsnm{Spirtes},~\bfnm{Peter}\binits{P.}},
\bauthor{\bsnm{Meek},~\bfnm{Christopher}\binits{C.}} \AND
\bauthor{\bsnm{Richardson},~\bfnm{Thomas}\binits{T.}}
(\byear{1999}).
\btitle{An algorithm for causal inference in the presence of latent variables
and selection bias}.
In \bbooktitle{Computation, Causation, and Discovery}
\bpages{211--252}.
\bpublisher{AAAI Press}, \baddress{Menlo Park, CA}.
\bid{mr={1689972}}
\bptok{imsref}%
\end{bincollection}
%
\endbibitem

\bibitem{VermaPearl90}
%
\begin{binproceedings}[author]
\bauthor{\bsnm{Verma},~\bfnm{Thomas}\binits{T.}} \AND
\bauthor{\bsnm{Pearl},~\bfnm{Judea}\binits{J.}}
(\byear{1990}).
\btitle{Equivalence and synthesis of causal models}.
In \bbooktitle{Proceedings of the Sixth Annual Conference on
Uncertainty in
Artificial Intelligence}
\bpages{255--270}.
\bpublisher{Elsevier}, \baddress{New York}.
\bptok{imsref}%
\end{binproceedings}
%
\endbibitem

\bibitem{Zhang08-causal-reasoning-ancestral-graphs}
%
\begin{barticle}[mr]
\bauthor{\bsnm{Zhang},~\bfnm{Jiji}\binits{J.}}
(\byear{2008}).
\btitle{Causal reasoning with ancestral graphs}.
\bjournal{J. Mach. Learn. Res.}
\bvolume{9}
\bpages{1437--1474}.
\bid{issn={1532-4435}, mr={2426048}}
\bptok{imsref}%
\end{barticle}
%
\endbibitem

\bibitem{Zhang08-orientation-rules}
%
\begin{barticle}[mr]
\bauthor{\bsnm{Zhang},~\bfnm{Jiji}\binits{J.}}
(\byear{2008}).
\btitle{On the completeness of orientation rules for causal discovery
in the
presence of latent confounders and selection bias}.
\bjournal{Artificial Intelligence}
\bvolume{172}
\bpages{1873--1896}.
\bid{doi={10.1016/j.artint.2008.08.001}, issn={0004-3702}, mr={2459793}}
\bptok{imsref}%
\end{barticle}
%
\endbibitem

\bibitem{ZhaoYu06}
%
\begin{barticle}[mr]
\bauthor{\bsnm{Zhao},~\bfnm{Peng}\binits{P.}} \AND
\bauthor{\bsnm{Yu},~\bfnm{Bin}\binits{B.}}
(\byear{2006}).
\btitle{On model selection consistency of {L}asso}.
\bjournal{J. Mach. Learn. Res.}
\bvolume{7}
\bpages{2541--2563}.
\bid{issn={1532-4435}, mr={2274449}}
\bptok{imsref}%
\end{barticle}
%
\endbibitem

\end{thebibliography}
\end{document}